\documentclass{emulateapj}

\def\arcdeg{\hbox{$^\circ$}}

\def\arcsec{\hbox{$^{\prime\prime}$}}
\def\deg2{\hbox{$\rm deg^{2}$}}

\def\lsim{\mathrel{\rlap{\lower4pt\hbox{\hskip1pt$\sim$}}\raise1pt\hbox{$<$}}}                
\def\gsim{\mathrel{\rlap{\lower4pt\hbox{\hskip1pt$\sim$}}\raise1pt\hbox{$>$}}}                

\renewcommand{\thefootnote}{\fnsymbol{footnote}}

\lefthead{Drake et~al.}
\righthead{Halo RR Lyrae}

\begin{document}
\title{Probing the Outer Galactic halo with RR Lyrae from the Catalina Surveys}

\author{
A.J.~Drake\altaffilmark{1}, M.~Catelan\altaffilmark{2,3}, S.G.~Djorgovski\altaffilmark{1}, 
G.~Torrealba\altaffilmark{2}, M.J.~Graham\altaffilmark{1}, V. Belokurov\altaffilmark{4},\\
S. E. Koposov\altaffilmark{4}, A.~Mahabal\altaffilmark{1}, J.L.~Prieto\altaffilmark{5}, 
C.~Donalek\altaffilmark{1}, R.~Williams\altaffilmark{1}, S.~Larson\altaffilmark{6},\\ 
E.~Christensen\altaffilmark{6} and E.~Beshore\altaffilmark{6}
}

\altaffiltext{1}{California Institute of Technology, 1200 E. California Blvd, CA 91225, USA}
\altaffiltext{2}{Pontificia Universidad Cat\'olica de Chile, Departamento de Astronom\'ia y Astrof\'isica, 
Facultad de F\'{i}sica, Av. Vicu\~na Mackena 4860, 782-0436 Macul, Santiago, Chile}
\altaffiltext{3}{The Milky Way Millennium Nucleus, Av. Vicu\~{n}a Mackenna 4860, 782-0436 Macul, 
Santiago, Chile}
\altaffiltext{4}{Institute of Astronomy, Madingley Road, Cambridge CB3 0HA, UK}
\altaffiltext{5}{Department of Astronomy, Princeton University, 4 Ivy Ln, Princeton, NJ 08544}
\altaffiltext{6}{The University of Arizona, Department of Planetary Sciences,  Lunar and Planetary Laboratory, 
1629 E. University Blvd, Tucson AZ 85721, USA}

\begin{abstract} 
  
  We present the analysis of 12227 type-ab RR Lyrae found among the 200 million public lightcurves in the Catalina
  Surveys Data Release 1 (CSDR1\footnote{http://crts.caltech.edu/}). These stars span the largest volume of the Milky Way
  ever surveyed with RR Lyrae, covering $\sim$ 20,000 square degrees of the sky ($0\arcdeg < \alpha < 360\arcdeg \!,$
  $-22\arcdeg < \delta < 65\arcdeg$) to heliocentric distances of up to 60kpc.  Each of the RR Lyrae are observed
  between 60 and 419 times over a six-year period.  Using period finding and Fourier fitting techniques we determine
  periods and apparent magnitudes for each source. We find that the periods at generally accurate to $\sigma = 0.002$\%
  by comparison with 2842 previously known RR Lyrae and 100 RR Lyrae observed in overlapping survey fields, We
  photometrically calibrate the light curves using 445 Landolt standard stars and show that the resulting magnitudes are
  accurate to $\sim 0.05$ mags using SDSS data for $\sim 1000$ blue horizontal branch stars and 7788 of the RR Lyrae.
  By combining Catalina photometry with SDSS spectroscopy, we analyze the radial velocity and metallicity distributions
  for $> 1500$ of the RR Lyrae. Using the accurate distances derived for the RR Lyrae, we show the paths of the
  Sagittarius tidal streams crossing the sky at heliocentric distances from $20$ to $60$ kpc.By selecting samples of
  Galactic halo RR Lyrae, we compare their velocity, metallicity, and distance with predictions from a recent detailed
  N-body model of the Sagittarius system. We find that there are some significant differences between the distances 
  and structures predicted and our observations.

\end{abstract}

\section{Introduction}

The study of the formation of the Milky Way is fundamental to the understanding of our galactic environment as well as
that of galaxies in general. The past competing ideas of halo formation through monolithic collapse (Eggen et al.~1962),
and through the accretion of protogalactic fragments (Searle \& Zinn 1978), have largely been replaced by a combination
of the two scenarios within the theory of hierarchical structure formation (e.g., Freeman \& Bland-Hawthorn 2002).

The study of Galactic structure has continued to flourish in recent years with the discoveries of numerous tidal streams
and dwarf galaxies within the Galactic halo (e.g., Majewski et al. 2003; Belokurov et al. 2006; Newberg et al. 2002).
The most well-studied of these structures is the accretion of the Sagittarius dwarf galaxy (Sgr, Ibata et al. 1994). To
date the Sgr stream has been traced on large scales using blue horizontal branch (BHB) stars (Newberg et al. 2003;
Belokurov et al. 2006) and M-giants (Majewski et al. 2003) with portions of the stream being studied using RR Lyrae
(RRL, Vivas \& Zinn 2006; Miceli et al.~2008; Sesar et al.~2010).  In addition to the Sgr stream, evidence has come for
a Virgo stellar stream (VSS; Vivas \& Zinn 2006; Vivas et al. 2008) using RRLs and a Virgo overdensity (VOD, Newberg et
al. 2002) from F-type main-sequence stars (Newberg et al 2007).  An overdensity in Pisces was reported by Sesar et
al.~(2007) and confirmed by Kollmeier et al.~(2009).  A Monoceros stream has also been discovered (Newberg et al. 2002;
Majewski et al 2003) that may be due to the disruption of the putative Canis Major dwarf (Casetti-Dinescu et al. 2006).
Although the existence of this structure remains uncertain (Momany et al. 2004, Mateu et al. 2009)
In addition, a Cetus stream has been discovered in the south (Newberg et al. 2009; Koposov et al. 2012) and Belokurov et
al.~(2007) also note the presence of an overdensity of BHB stars dubbed the Hercules-Aquila Cloud.

Although the number of streams and structures found in the outer Galactic halo (galactocentric distances $>15$ kpc)
has significantly increased in the past ten years, numbers fall far short of the hundreds predicted by 
$\Lambda$CDM models of hierarchical structure formation (e.g. Bullock et al.~2001; Freeman \& Bland-Hawthorn 2002). 
This ``missing satellite'' problem (e.g. Bullock et al.~2001) continues to be important to our understanding
of galaxy formation and requires us to probe the Galactic halo to distances well beyond $20$ kpc.

RR Lyrae are fundamental distance probes that can be used to trace the history of galaxy formation (e.g., Catelan 2009,
and references therein).  To date a few tens of thousands of RRL are known in dense regions near the Galactic bulge,
where the Sgr dwarf galaxy is located. Similar numbers of RRL are known also in the Magellanic Clouds, again thanks to
microlensing surveys towards dense stellar fields (Soszy\'nski et al. 2009, Pietrukowicz et al. 2012). However, the
Galactic halo itself has only been probed with confirmed RRL over a few thousand square degrees to heliocentric 
distances of $\sim 30$ to 100 kpc (eg., Vivas et al.~2004; Keller et al.~2008; Miceli et al.~2008; Watkins et al.~2009; 
Sesar et al.~2010). In this paper we outline our search, discovery and calibration of the RRL to $\sim$ 50 kpc. We
then undertake a preliminary analysis of the structures uncovered.

\section{Observational Data}

The Catalina Sky Survey\footnote{http://www.lpl.arizona.edu/css/} began in 2004 and uses three telescopes to cover the
sky between declination $\delta = -75$ and +65 degrees in order to discover Near-Earth Objects (NEOs) and Potential
Hazardous Asteroids (PHAs).  Each of the survey telescopes are run as separate sub-surveys.  These consist of the
Catalina Schmidt Survey (CSS) and the Mount Lemmon Survey (MLS) in Tucson Arizona, and the Siding Spring Survey (SSS) in
Siding Spring Australia.  In general each telescope avoids the Galactic plane region by between 10 and 15 degrees due to
reduced source recovery in crowded stellar regions.  All images are taken unfiltered to maximize throughput. Photometry
of all images is carried out using the aperture photometry program SExtractor (Bertin \& Arnouts 1996). In addition to
asteroids, all the Catalina data is analyzed for transient sources by the Catalina Real-time Transient Survey (CRTS,
Drake et al.~2009; Djorgovski et al 2011).

In this paper we concentrate on the data taken by the Catalina Survey Schmidt 0.7m telescope (CSS) between April 2005
and June 2011. These data cover 20,155 square degrees in the region $0\arcdeg < \alpha < 360 \arcdeg$ and $-22\arcdeg <
\delta < 65 \arcdeg$. For this CSS telescope each image from the $4{\rm k} \times 4{\rm k}$ Catalina CCD camera covers 8
sq. degrees on the sky.  All these archival observations analyzed in this work were taken during spans of 21 nights per
lunation in sets of four images separated by 10 minutes.  The exposure times are typically 30 seconds and reach objects
to $V = 20$ mag, depending on seeing and sky brightness. The distribution of observations in the CSS fields is 
given in Figure \ref{NfNo}.

\section{Calibration}

In order to use RRL as probes for distances it is necessary to accurately calibrate the observed magnitudes
to a standard system. All observations are transformed to Johnson $V$ based on 50-100 stars selected as G-type stars using
2MASS (Skrutskie et al.~2006) colours.  For bright stars, this photometry provides repeated photometry accurate to $\sim
0.05$ (Larson et al. 2003).  However, as the photometry is unfiltered there are significant variations with object colour.
The first step to determining accurate distance is calibration of the colour terms required and thus place the
photometry on a standard system.

The Landolt (2007) UBVRI standard star catalog provides 109 stars centered near declination $-50\arcdeg$ in the magnitude
range $10.4 < V < 15.5$ and in the color index range $-0.33 < (B-V) < 1.66$, while Landolt (2009) provides
a catalog of 202 standard stars along the celestial equator in the magnitude range $8.90 < V < 16.30$, and the color
index range $-0.35 < (B - V) < 2.30$, along with 393 standard stars from previous standard star catalogs.

For photometric calibration we combine observations taken with all three Catalina Sky Survey telescopes since no
difference was found between these systems. This is not surprising since each telescope specifically uses the same type
$4{\rm k} \times 4{\rm k}$ CCD camera and observations with all three telescopes are calibrated using the same software
pipeline. In total there are 445 Catalina lightcurves matching Landolt standards.  On average each standard is measured
134 times. To reduce sources of error we first determined the variability index of each lightcurve. We remove a handful
of stars that appeared to exhibit significant variability. As noted by Landolt (2009) this catalog includes a small
number of known variable stars.

In order to compare CSS $V$ magnitudes to Landolt values we first determined median magnitudes for each lightcurve and
calculated the difference from the standard value. In Figure \ref{Landolt}, we plot the difference between Landolt
standards and transformed median CSS magnitudes. The high degree of scatter is due to the clear difference between the
transformed unfiltered observations and filtered observations. The Landolt dataset contains values for U, B, V, R and I
filters. To better calibrate the photometry we fit the differences with the various possible colour terms. The colour
transformations were clearly nonlinear.  We find the follow transformations:

\begin{eqnarray}\label{tran}
\rm V = V_{CSS} + 0.31 \times (B-V)^2 + 0.04,\\
\rm V = V_{CSS} + 0.91 \times (V-R)^2 + 0.04,\\
\rm V = V_{CSS} + 1.07 \times (V-I)^2 + 0.04.
\end{eqnarray}
\vspace{0.1cm}
In Figure \ref{Landolt}, we also plot the difference in magnitudes after applying this calibration. The 
dispersion in the fits to these transformations are 0.059, 0.056 and 0.063 magnitudes, respectively
for $V < 16$.

The average B-V colour of RRL is about 0.3 mag with stars varying between about 0.1 and 0.5 as they pulsate (Nemec
2004).  For most of our sample we have no colour information so we adopt the average colour. From equation (1), this
leads to a correction of 0.028 magnitudes. From the transformations the range of possible colours gives rise to a
maximum uncertainty of $\sim 0.07$ mags in $V$. Combining this with the photometric uncertainty, we expect a dispersion of
$\sigma = 0.09$ mags in our RRL photometry.  However, based on random phase SDSS photometry, we will show that the RRL
in our sample are strongly concentrated in a colour index range of 0.1 magnitudes. Further tests of the importance of
RRL colour variations in CSS data were carried out by Torrealba et al.~(2012) and found to be unimportant.

To determine the photometric accuracy of the calibration at fainter magnitudes than Landolt standards we selected the
sample of 1170 BHB stars of Sirko et al.~(2004). These stars have similar ages, masses and $(u-g)$ colours to RRL stars.
However, they have significantly different $(g-r)$ colours (centered near -0.2, compared to +0.2 for the RRL).  We matched
the BHB source locations with CSS objects and removed 93 sources that matched candidate or known variable stars. As we
want to determine robust estimates, we also removed objects with fewer than 20 CSS observations.  The average number of
measurements for the remaining 1026 BHB stars was 223.  We transformed the SDSS DR8 photometry for the source to $V$ using
the Lupton (2005) transformation equations.  For each source we determined median CSS V-magnitudes using equation
\ref{tran} and $B-V$ colours from SDSS data.  In Figure \ref{BHB}, we plot the difference between $V$ magnitudes derived 
from SDSS and CSS and by binning the difference in one magnitude bins we also show how the scatter increases with 
decreasing source brightness.

Of the 1026 BHB stars, 14 had offsets $>0.3$ magnitudes. Nine of these objects were found to be blended in the CSS
photometry (but not in SDSS photometry). For the remaining objects we find an average difference of $0.0065$ magnitudes
and $\sigma = 0.065$. As expected, the level of variation increases with decreasing brightness. Considering
that these stars are much fainter than the Landolt standards the level of agreement is very good.

\section{Sample Selection}

The Catalina Sky Survey data release 1 (CSDR1) covers 198 million discrete sources ranging in $V$ from 12 to 20 with an
average of 250 observations per location. 
In order to discover the RRL among these sources we first calculate the Welch-Stetson variability index $I_{WS}$ (Welch
\& Stetson 1993) for every source.  Based on initial investigation we selected sources with $I_{WS} > 0.6$ as possible
variables.  For sources brighter than $V=13.25$ we set a higher variability threshold due to saturation effects.  We also
only selected lightcurves with more than 40 points. This process returned 8.7 million potential variable sources, or
4.4\% of the sources.  Every candidate variable source was then checked for periodicity using the Lomb-Scargle (LS, Lomb
1976; Scargle 1982) periodogram analysis. This method was chosen since it was found to take approximately a second per
lightcurve, compared to between 10 seconds and a few minutes for other techniques.  Periodic sources were selected based on
LS peak significance statistic $p_0$. This value represents the probability that the observed signal was observed purely
due to chance fluctuations. However, we note that care must be taken when interpreting these values
(Schwarzenberg-Czerny 1998). Objects with very small values, $p_0 < 10^{-7}$, were chosen as good candidate periodic
variables after the inspection of a few hundred sources phased to their best LS periods.  Based on the CSS lightcurves
of known periodic variables, Graham et al.~(2012, in prep.) found that this technique gives the correct periods for
$\approx$ 83\% of known RRab with this sampling. An additional 13\% of the RRL were detected as significantly periodic, yet the period
did not match the known period.

375,000 of the variable sources were found to exhibit periodicity at our significance level. However, a very large
fraction of the periods were found to be spurious detections near 0.5 and 1 days. These periods are purely due to the
observing cadence of CSS. Upon close examination of the period distribution we removed all candidates with periods in
the ranges $0.497 < P < 0.501$ or $0.994 < P< 1.0035$ days.  Additional period aliases were found within individual
fields.  In these cases systematics were found to lead to the detection of many sources with very similar periods. We
also removed periodic variable candidates where three or more sources within a given field had the same periods to 
$<$ 0.2\%.  These cuts are expected to remove a very small fraction of the RRL with periods within these ranges.

To obtain a clean sample of RRL appropriate for distance determinations we decided to use only type-ab RRL
(RRab), since c-type (RRc) and d-type (RRd) lightcurves are often very similar to W~UMa type eclipsing binary
lightcurves that occur in the same period region.  For example, some past surveys have misidentified W~UMa 
sources as RRc's (Kinman \& Brown 2010). W~UMa stars are more common than RRc variables, so even though most
can be distinguished in well-sampled data, including RRc's is likely to lead to some contamination.

Among the periodic sources we initially selected objects with periods between 0.36 and 1.4 days
to conservatively include all RRab found at their true period, as well as many of those found at aliases of their period
in the LS analysis.  A total of 23346 objects were found in this period range.  From the phased lightcurves a large
number of these sources were clearly eclipsing binaries of all types, as well as RRc's found at aliases of their true
periods.

To recover the 13\% of sources that we expect to be detected at an incorrect period, and to determine more accurate periods 
among the periodic candidates, we determine the ten best periods for each source using the Analysis of Variance program (AoV)
(Schwarzenberg-Czerny 1989). The AoV program was run in two stages which each provided five period. Firstly the software
was run in the normal {\em AoV-mode} and then with the multi-harmonic method (Schwarzenberg-Czerny 1996).  
We compare these periods to the five best values found using the LS technique, giving us a total of 15 test periods.
As there were many cases where the most significant period found by each method did not match that given by others, 
further analysis was undertaken.

To select the correct period for each object we follow the Adaptive Fourier Decomposition (AFD) method (Torrealba et al.
2012, in prep.). Here the phased lightcurves were fit with an increasing series of Fourier harmonics using a weighted
least-squares technique.  The order of the harmonic fit is chosen by determining whether the improvement in the
observed reduced $\chi^2$ with additional of higher-order terms, is statistically significant based on the 
statistical F-test. This is evaluated by determining the likelihood of observing the improvement in reduced $\chi^2$
given the number of parameters and avoids over fitting the data with a single high-order series. 
In addition, since type RRab require more Fourier terms to fit than than most eclipsing binaries and RRc, the 
fit order also provides additional discrimination between variable star types.

We produced phase folded lightcurves using the 15 most significant periods from AoV and LS. These were fit to select the
best period based on their reduced $\chi^2$ values.  In addition, to minimize the influence of bad data, each object is
refit successively after removing outliers $3 \sigma$ from the original fits. For RRab selection we remove sources where
the best fit to the phased light curves is sinusoidal.  This is done by only selecting objects where the best Fourier
fit is order three and above.  Objects with large reduced $\chi^2$ values at all periods were removed from the candidate
list.

To further separate the RRab from W~UMa variables we select only objects where the best fit period among the 15
candidates is between 0.43 and 0.95 days.  For the remaining set we apply the M-test (Kinemuchi et al.~2006).  This test
statistic measures the fraction of time that an object spends below the mean magnitude.  Lightcurves with $M > 0.502$
are selected as RRab since these sources spend most of the time above their average magnitude.  In Figure \ref{mtest},
we show the best fit periods of all the sources and their M-test values. The dashed-lines show the region where RRab
sources are selected. Lightcurves with $M = 0.5$ are sinusoidal W~UMa and RRc variables. The eclipsing binaries are
mainly concentrated at short periods near 0.2 days since these sources are very often found at half their true period.
The RRc's are seen near 0.4 days but, as already explained, they are likely contaminated by some eclipsing binaries. The
cutoff seen near 0.35 days and the slight gap near 0.5 days are due to the initial period selection limits. Many objects
are seen near 1 day due to the daily sampling-rate alias.  Of the initial selection, 12471 are selected as RRab stars.

In Figure \ref{PerAmp}, we plot the period-amplitude distribution of CSS RRab's. In the right panel of this figure we
also present a Hess (relative density) diagram for this same data.  This figure shows that almost all of the RRab's lie
near the Oosterholf type-I (OoI) period-amplitude sequence. However, the amplitudes are slightly smaller than predicted
since we have assumed average $B-V$ colour, whereas RRab's vary in colour with phase (Hardie 1955). From equation (1), 
we find that a $B-V$ colour variation, between $B-V=0.1$ at maximum and $B-V=0.5$ at minimum, would lead to a 0.08 mag
increase in the $V$ amplitude. By comparison with the RRab $V$ amplitudes of Zorotovic et al.~(2010) find that the CSS 
RRab amplitudes uniformly underestimated by 0.15 mags and correct for this factor.

A sequence of RRab's are seen at longer periods due primarily to lower-metalicity Oosterholf type-II (OoII) RRab's
(e.g., Smith et al.~2011).  However, the fraction of of Oosterholf-II stars is far smaller than observed for a sample of
1455 RRab nearby ($\rm d < 4 kpc$) observed by Szczygiel et al.~(2009).  The figure exhibits the presence of a gap in
the distribution near 0.5 days due our period selection where we removed periods near sampling aliases.  Based on the
number of RRab's with slightly shorter and slightly longer periods, we estimate the number of stars in this range to be
around 160 stars, or $\sim 1.3\%$ of the total.

Using the OoI period-amplitude relation defined by Zorotovic et al.~(2010) we determine period-shifts for each RRab as
shown in Figure~\ref{PerAmpM}. The result is remarkably similar to the one shown in Figure~20 of Miceli et al.~(2008),
thus being consistent with OoI and OoII components being present in our data as well. The fit with two Gaussian
components that is shown in the figure possesses a correlation coefficient $r = 0.985$ and a standard error of 25.4.  We
also fitted a skew-normal distribution to the data, based on equation~(3) of Azzalini (1985). The result is shown in
Figure~\ref{PerAmpM}. This fit is noticeably worse than the two-Gaussian fit, with $r = 0.955$ and a standard error of
the estimate of 43.3. This confirms that, as in the case of Miceli et al.~(2008), our distribution is also comprised of
two separate components, which are naturally interpreted as OoI and OoII. However, the OoII component is clearly smaller
in our case; the two-Gaussian fit implies that around 76\% of our stars belong to the OoI, and 24\% to the OoII
population.  There is, in addition, a clear excess of stars towards negative period shifts, which can be plausibly
ascribed to the presence of the Blazhko effect as well as RRab's that are blended with other sources.

We examined the phased lightcurves of all the objects outside the region bounded by the dashed-lines in Figure
\ref{PerAmp}. Of the 439 RRab candidates in these regions, 140 were discovered not to be RRab and removed.  Most of the objects
removed were variable stars near the CSS saturation limit, $V \sim 12.5$.  Many of the objects with unexpected
amplitudes for an RRab were found to be blended sources. For close blends the additional flux tends to reduce the
observed amplitude. For sources with slightly larger source separations the amplitude can actually increase slightly. In
such cases the flux from the two separate sources are detected individually at minimum. As the RRab's flux increases,
the nearby source become merged with the RRab flux.

The average number of photometric measurements for the RRab candidates is 219, with the poorest sampled having 60 and
the best sampled having 416 measurements.  In Table 1, for each source we present the locations, average $V$ magnitudes,
periods, amplitudes, number of photometry measurements, distances and extinction.  In Figures \ref{AitEq} \&
\ref{AitEqlb} we present the locations of the RRab discovered in CSS survey fields.

In Figure \ref{RRLC}, we plot representative examples of RRab lightcurves spanning the range of discoveries from $V=12.5$
to 19.5. Each lightcurve has been folded with the period we discovered. The figure shows that the brightest RRab, near
mag 12.5, show significant saturation effects.  This will affect the fits for these sources. However, there are only 62
RRab in our sample brighter than $V=12.5$. A small fraction of the points shown in other lightcurves are clearly outliers
caused by image artifacts and poor seeing. As outlined above these points are removed during the period finding and
Fourier fitting process.

Of the remaining objects, 100 sources are RRab's observed in the overlap regions between fields. These sources provide a
useful test of the photometric calibration between CSS fields and the period determinations. In Figure \ref{Diff}, we
plot the difference between the average fit magnitudes and periods of the RRab's overlapping between fields.  This
suggests that uncertainties in the photometric calibration between fields are generally $< 0.1$~mag in agreement with
the comparison with Landolt stars. We note that since overlapping objects are located on the edges of the fields, where
the photometry is poorest, on average the photometry should be slightly better. In addition, we can see that the period
determinations are in excellent agreement even though the total number of observations, and observation dates, vary
between adjacent fields.  For an RRab with a 0.6 day period, a 0.004\% uncertainty corresponds to an uncertainty of only 2
seconds. Of the 100 objects, none differed by more than $0.007\%$ in period, or more than 0.12 mag in $V$.

\newpage
\section{Comparison with RR Lyrae from past Surveys}

To compare the CSS RRab parameters with those of known RRab's, we downloaded all the objects marked as variable sources
in the Simbad database. This consisted of 41765 objects. We also downloaded the International Variable Star Index (VSX,
Jan 2011 edition; Watson et al.~2006) dataset and extracted all the sources marked as RRL. This consisted of 24124
objects. A small number of the VSX objects are duplicates based on their positions. Simbad and VSX data significantly
overlap, yet both contain some unique sources. In the bulk of cases where the two sets match, the VSX dataset is
superior since it provides the periods for the RRL.

We matched the RRab with the VSX and Simbad datasets and found 2136 matches to known Simbad variable sources and 2753
matches to VSX sources. In order to account for significant astrometric uncertainties in some of the older sources we
used a large 10 arcsecond matching radius. From the combined datasets we find matches to 2842 known sources. This is a
small fraction of the total number of known RRL. Most have been found by microlensing surveys which have almost
exclusively covered the Galactic bulge and the Magellanic Clouds (e.g., Soszy\'nski et al. 2009, Pietrukowicz et al. 2012).
Additional large numbers of sources come from globular clusters near the Galactic plane that are not covered by CSS.  Of the VSX
RRab matches, 2727 objects have recorded periods. In Figure \ref{Hist}, we plot the magnitude distribution
of the CSS RRL compared to that of these previously known sources.

In Figure \ref{DiffVSX}, we plot the percentage difference between VSX periods and those derived from CSS data for the
matching previously known RRL. As with the 100 overlapping CSS RRab's the scatter in periods is generally $\sigma \sim
0.002\%$.  However, 397 of the RRL have period differences of $>0.01\%$. Objects with this level of uncertainty
would have a phase error of 0.1 over 1000 cycles (or time spans from one to three years for RRab).  In 49 cases
the difference in period was greater than $1\%$. We checked the phased lightcurves for the 397 objects and found only
two objects where the CSS period was incorrect. Four objects had similar lightcurve reduced $\chi^2$ with both CSS and
VSX periods. However, three of these were apparent aliases, since they were noted with periods $< 0.41$ days.
The remaining RRab had a $0.012\%$ period difference with both periods being equally likely.  Of the objects with
apparently incorrect VSX-periods, 89 were from the NSVS sample (Kinemuchi et al. 2006), and 61 were from the LONEOS-I
sample (Miceli et al.~2008).  Since only two of the 2675 matching sources had clearly incorrect CSS periods we have high
confidence in the periods derived here.  

Figure \ref{Hist} shows that most the VSX-CSS matching sources are on average brighter than RRab in the full sample.
This suggests that the matched objects will, on average, be better sampled and have a higher signal-to-noise ratio 
than the sources in general. It is likely that the faint CSS RRab's have less accurate periods than the
bright ones.

In order to get an idea of our detection completeness we extracted lightcurves for all the VSX sources marked as
possible RRab's within our survey region limits. We found CSS matches to 4144 VSX RRab sources. In a number of cases
there were multiple VSX objects at the same location to within a few arc seconds. These sources are very likely duplicates.

Of the 1328 unique VSX RRab with periods that were missed in our RRab selection process, 298 objects were not selected by
our Welch-Stetson variability of $I_{WS} > 0.6$.  Of these, 97 had either $V < 12.5$, and were either saturated, or had
$V > 19$ and were too faint for us to detect their variability. Most of the remaining VSX objects not in among our
candidate variables were poorly sampled. Many had fewer than the 40 measurements required for initial selection.
Inspection of $\sim 100$ of the CSS lightcurves for these sources showed that a dozen were not variable, and a 
couple of dozen were blended with nearby stars.

Of the remaining 1030 VSX sources selected as variable in CSS data, 627 had LS periods with significance below our 
threshold (that is $p_0 > 10^{-7}$). Of these low periodic significance sources, 290 had periods outside our 
$0.36 < P < 1.4$ day pre-selection window, and 27 have VSX periods less than 0.4 days.

Inspection of the CSS photometry for the 403 remaining VSX sources with periods and significance within our selection range,
showed that 52 were not RRab, but RRc's and other types of variables. An additional 52 appear to be Blazkho RR Lyraes
(Blazhko 1907). These Blazhko RRab candidates were likely missed due to poor Fourier fits in the presence of phase 
variations. Among the remaining 299 candidates, 66 were poorly sampled, and 233 had noisy CSS light curves due 
to blending and saturation effects.

Overall we find $\sim30\%$ of the genuine RRab in the VSX dataset were missed in our selection 
because of their brightness (2\%), blending (1\%), poor or noisy lightcurve sampling (8\%),
period variations (1\%), or inaccurate LS periods (17\%).

\subsection{Completeness}

In order to better understand the detection completeness of our RRab sample we decided to simulate the detection of CSS RRab's
from lightcurves through to variability selection and processing with AFD software. This estimation process
requires understanding the different sampling effects and variation in uncertainties between fields, as well
as reproduction of realistic RRab lightcurves and underlying period distribution.

The CSS data set analysed here contains 2454 separate fields. As shown in Figure \ref{NfNo}, there is significant 
range in the number of observations per field. The distribution of observation density on the sky is given on the 
Catalina data release website\footnote{http://catalinadata.org}. In our simulations we selected $\sim 5\%$ (134) of 
the fields.
The least sampled among these test fields had 35 observations while most sampled had 256. For each of the 134 test fields
we measured the average magnitude and scatter in brightness for each source.  We also determined the detection
completeness as a function of magnitude for each image based on comparison with deeper coadded images. As systematic
uncertainties are likely to vary between fields, we used the average source magnitudes and uncertainties to determine the
scatter as a function of magnitude for each field. Since $\sim 5\%$ of the photometry was found to contain outliers we
model the error distribution using two separate Gaussians with varying standard deviations. One reflects the $95\%$ of good
data and the the remaining points model outliers.

To simulate realistic lightcurves we selected 1010 high signal-to-noise CSS RRab's with average magnitudes $V < 16.5$,
Fourier-fit reduced $\chi^2 < 1.5$, and more than 100 observations.  The fits to these sources serve as templates. As
the detection completeness depends on RRL period it is necessary to select an underlying period distribution. Cseresnjes
(2001) provides a sample of 3700 RRLs from a mixture of Sagittarius dwarf and Milky Way sources. We used this
distribution to select periods for our test sources. The number of test sources in each field was chosen to
exponentially increase with decreasing brightness so that many faint sources would properly sample the detection
sensitivity for faint sources.  Once a period is selected we find the closest match among the 1010 templates and combine
this with the uncertainties observed for each field and brightness. We generated $\sim 100,000$ artificial RRab
lightcurves for sources with magnitudes from $V=12.5$ to 20.5.

The artificial RRab lightcurves were all run through the same variability and period selection process as the real data. Of the
100,000 simulated objects, 15,271 were detected as variable sources and thus had their LS periods determined.  Of these
sources 11,543 were found to have periods within the range selected for RRab's and among these 10,483 (90\%) were
ultimately selected as RRab via the AFD software.

In Figure \ref{Compl}, we show the distribution of recovered artificial RRab's. The error bars show Poisson uncertainties
based on the number of detected sources. The figure suggests that real RRab's will be missed at all brightness levels.
This result is in agreement with our comparison to VSX data.  Also, few of the brightest sources are recovered because
saturation effects cause large uncertainties.  This figure combines the poorly sampled fields with well sampled
fields. On average the artificial lightcurves have fewer observations than the observed distribution and thus
underestimate the average number of recovered sources. However, the plot clearly shows that many of the faint objects
that are selected as variable sources are ultimately not recovered as RRab's. 
  This result suggests that implying additional period finding searches on the millions of candidate variables
may well lead to additional RRab discoveries. In particular, as the largest difference in recovery is at 
faint magnitudes many distant RRab's may be recovered. However, period recovery at such low signal-to-noise may 
be difficult. 

To investigate the dependence of completeness on the number of observations in a field, we combined the 134 fields into
four groups.  Fields observed less than 100 times, fields observed between 100 and 130 times, fields observed 130 to 180
times and fields observed more than 180 times. The completeness results for these groups are also shown in Figure
\ref{Compl}.  The figure clearly demonstrates the significant affect that increasing numbers of observations have.  Much
of the reason for this difference is mainly is the decrease in photometric sensitivity with magnitude. For example, a
19th magnitude source is detected in less than half of the observations of a field. Thus the number of points within the artificial
lightcurves of faint sources is far fewer than the total number of observations.  The average number observations per
field best matches the top curve we present, suggesting that $\sim 70\%$ of the bright RRab's are recovered (in good
agreement with the analysis above). We note that our completeness for small numbers of observations is in marked
contrast to the results Miceli et al.~(2008) based on between 28 and 50 epochs of LONEOS data.  However, these authors
detected RRLs using a template fitting method. This is likely to be much more sensitive with smaller numbers of
observations and may also improve the recovery of faint RRab's.

\section{Additional Information from SDSS}

For most CSS RRab's above declination $\sim -2\arcdeg$ it is possible to check the accuracy of the transformed CSS
photometry using SDSS data.  Within the stripe-82 region the SDSS data has recently been shown to have photometric
uncertainties of 1\% or less (Ivezic et al. 2007). The SDSS data also reaches objects significantly deeper than our RRL
sample and provides spectroscopic information for very large numbers of sources in our RRab catalog.

\subsection{SDSS Photometry}

To carry out a photometric comparison we matched the locations of our RRL using the SDSS cross-match service and a $3
\arcsec$ radius. We select the nearest source within this region as the best match. Of the 12331 RRL
we find 8746 sources in SDSS DR8 that match our RRab's.  SDSS data saturates in r,i and z for stars brighter than 
$\sim 14.5$ and in u-band at magnitude 16 (Chonis \& Gaskell 2007).
After removing the matching objects above the saturation limit we find 7788 sources with SDSS photometry. 
We correct for the extinction of SDSS photometry using the Schlegel, Finkbeiner \& Davis (1998)
reddening maps and coefficients.
In Figure \ref{sdss}, we plot the extinction-corrected g vs $(u-g)$ and $(u-g)$ vs $(g-r)$ photometry for the RRab's, and in Figure
\ref{sdss2}, we plot the $(g-r)$ vs $(r-i)$ and $(r-i)$ vs $(i-z)$ colours of the objects with SDSS photometry.  The bulk of the sources
are strongly clustered near $(g-r)= 0.25$ and $(r-i)=0.1$ with a scatter of $\sim 0.1$ magnitudes. This suggests that the uncertainty
in our absolute $V$ magnitudes based on assuming the average $B-V$ of RRL is generally $< 0.05$ mag.

Almost all of the RRab's lie within the SDSS colour region selected by Ivezic et al.~(2005). We inspected a number of
the outliers and it was clear that a small number of the RRab's were blended with other stars, or galaxies, when compared 
to higher-resolution SDSS images.  The flux from the additional source distorts the colour of the object, increases the
brightness, and typically reduces the observed amplitude of the variability.

To compare the SDSS magnitudes to those derived from our calibration with Landolt standards, we first use the
photometric transformations of Ivezic et al (2007), which are themselves tied to Landolt standards via a large sample of
Stetson (2000, 2005) secondary standard stars.  Then following Sesar et al.~(2010, their eq. 13), we transform the
extinction-corrected SDSS photometry to $V$ magnitudes.  This is a slightly different transformation than used with the
BHB sample, since Ivezic et al.~(2007) note that their calibration is not appropriate for BHB star colours.

In Figure \ref{phase}, we plot the transformed values in comparison with the average CSS $V$ magnitudes from the Fourier
fits. The high degree of scatter ($\sigma=0.28$ mag) is due to the observations being taken at random phases.  Accurate
observation times were calculated for each SDSS measurement. For each object we took the average observation time of the
$g$ and $r$ measurements. The difference between these is only $\sim 4$ minutes and thus very small in comparison to the
RRab's periods. Using our Fourier fits we determine the phase of the RRab in the $V$-band lightcurves at the time SDSS
observed it, and then we use the fits to calculate the $V$ magnitude offset at this phase.  In Figure \ref{phase}, we plot
the magnitudes corrected for the SDSS phase offset. The overall reduction in scatter is evident.  After $3 \sigma$
clipping the clear outliers, the 1-sigma in magnitude difference is less than 0.12 mags and the average offset is 0.003
magnitudes. This dispersion is much larger than for the BHB sample.

Like the BHB sample, the dispersion in the phase-corrected and transformed SDSS magnitudes includes errors in the SDSS
photometry and uncertainties in the transformation from SDSS $g$ and $r$ to $V$ magnitudes. However, additional scatter
beyond the BHB sample is due to the uncertainties in the RRL periods (of order 0.002\%) and the resulting phase
corrections as many of the SDSS observations of these objects were made around 2001, while CSS observations were
generally taken from 2005.  For example, a 0.5-day period RRab undergoes $\sim 3000$ cycles in four years.  Over this
period a 0.002\% error in the period compounds to a phase offset of 0.06 ($\sim 43$ mins).  In RRab lightcurves an error
of $\sim 0.1$ in the phase before the peak can correspond to a 1 magnitude variation in brightness.  Taking this into account, the
average magnitudes derived from the CSS lightcurves should have accuracy similar to the BHB sample, with some additional
small uncertainty of up to 0.07 magnitudes due to the assumption of an average RRL colour (see \S 3).

To further investigate the source of differences in brightness we selected the 457 outlier RRab's with offsets $> 0.36$
mag ($3 \sigma$). From visual inspection it was clear that approximately 50 exhibited period variations.
The offsets due to period changes within these stars cannot be corrected without contemporaneous data.  Such
period changes are known to be common as Alcock et al.~(2000, 2003) found 10\% of the 6391 LMC RRab's stars they 
observed to exhibit period changes.  Many of the objects with offsets $< 0.36$ magnitudes will also exhibit period
variations, thus increasing the observed scatter.
A small fraction of the outliers were found to be due to sources that were blended in CSS images, yet were resolved in
the SDSS data.  Additionally, a small number of the objects were either matched to the wrong SDSS source or the SDSS
photometry was clearly spurious.  Of the remaining outliers, 68 had offsets due to slight errors in the period.

We interactively derived more accurate periods for 105 of the objects (including some likely Blazkho RRL).
The average difference between the original period and the new values was 0.0023\%.  In Figure \ref{phase}, we also plot
offsets for the original and improved periods for the 68 RRab's, as well as the offsets for period changing RRL. Of the
objects for which we obtained improved periods, eight remained with offsets $>0.36$ mags, yet we estimate the errors in
their new periods are $< 0.001\%$. Investigation of these sources revealed that they all had multiple epochs of SDSS
photometry. It is possible that the SDSS photometry in the database comes from the second epoch of images.
Additionally, four of the outlying sources were found not to be RRab's, reducing the total number to 12227.

\subsection{SDSS Optical Spectra}

The SDSS has released spectra of almost 1.6 million objects of which 460,000 are stellar objects (Abazajian et al.
2009).  Among these spectra are those from the SEGUE subproject which specifically consisted of 240,000 stars with
$14.0 < g < 20.3$ examined in order to study the structure of the Galaxy (Yanny et al. 2009). A second SEGUE survey
covering an additional 120,000 stars is yet to be released.  Each SDSS spectrum covers the 385-920nm range and has
resolution $R\sim2000$ with target $S/N\sim 25$. Matching our data set to SDSS DR8 spectra we found 1871 matches.

\subsubsection{Metallicities}

Among the 1871 SDSS spectra, 237 are multiple-observation RRab's having a total of 632 spectra. Some of these sources
have four or more observations. These sources provide an excellent way of determining the level of variation in the SDSS
spectra for RRab's. In particular, RRLs are well known to exhibit significant variation in radial velocity measurements
because of pulsation (eg. Lee 1991). For this reason, metallicities are traditional measured from the difference in spectral
type at minimum light measured from hydrogen lines compared to estimated from the $\rm C_{II}$ K line (Preston 1959).
Butler (1975) extended this method so that values could be obtained at phases other than minimum light. Other methods of
determining metallicity were also been devised use $\rm Ca_{II}$ K equivalent widths (Clementi et al.~1991).  Layden
(1994; figure 1) clearly shows the variation and overlap for hydrogen and calcium equivalent widths for RRL's of varying
metallicities and derived an iterative method for determining metallicity.

In contrast to these methods, the SDSS team applied 12 separate methods for determining [Fe/H] in DR8 via the SEGUE
pipeline (Lee et al. 2008, 2011).  None of the methods used by SDSS exactly matches that applied to RRab's.  The SDSS
measurements are calibrated based on the known metallicities of globular clusters that were specifically covered for
calibration purposes. The resulting values from the various methods are combined to provide an overall best value
(FEHADOP) along with an uncertainty.  

It is important to note that the spectra are taken irrespective of the phase of the sources.  When the RRab's are
far from minimum light variations in the spectra are usually not used in metallicity determination. Although For et
al.~(2011) suggests that spectra observed near near maximum light can also be useful for abundance studies.
SDSS spectra are also composites of multiple exposures.  In most cases these consist of observations from three back-to-back
900 second exposures. However, the composites can be spread over days (Bickerton et al.~2012). This means that the
RRab's have been be observed at many phases.

To investigate the affect of phase variations in composite SDSS spectra on adopted metallicities, we calculated the
variations in metallicity values for the objects observed on multiple nights.  In Figure \ref{FeMult}, we present a
histogram of the single-object [Fe/H] variations. A Gaussian fit to the distribution gives $\sigma =0.22$. The level of
variation is consistent with the range of metallicities observed over a pulsation cycle by For et al.~(2011). Thus
without considering any phase information the measurements are quite consistent.  This simple match between repeated
observations does not address possible systematic effects.  To accomplished this we must undertake comparisons with
known metallicities.

As globular clusters serve as metallicities standards, we searched for known RRab's within these associations. We matched
the Samus et al.~(2009) Catalog of Variable Stars in Globular Clusters (CVSGC) with spectra from SDSS DR8. We then
removed the non-RRab variables based on previous classifications of the underlying sources as well lightcurves extracted from
Catalina. Of the 52 variables with SDSS spectra, 26 were found to be RRab's, and among these 17 were from the well
studied cluster NGC 5272 (M3).

Numerous surveys have measured the metallicity of NGC 5272 (eg. Zinn \& West 1984; Armosky et al.~1994; 
Kraft et al.~1992; Sandstrom et al.~2001; Cohen \& Melendez 2005; Cacciari et al. 2005).
Based on the results from these surveys we find and average metallicity of $\rm [Fe/H] = -1.46$.
From the range of measurements we estimate the uncertainty to be $\sim 0.1$. 
In Figure \ref{M3}, we plot the metallicities and uncertainties for the RRab's provided by SDSS DR8.
After removing a single outlier (the most distant RRab at 7.4 half-light radii with $[Fe/H] = -1.76$), 
we find average $[Fe/H] = -1.38 \pm 0.10$.  Cacciari et al.~(2005) recently undertook an analysis of 
NGC 5272. Based on 45 RRab's they found $\rm [Fe/H] = -1.39\pm0.11$.  
The level of agreement in this case is clearly very good. However, NGC 5272 only 
serves as a single calibration point.

To investigate this further we searched the literature to find metallicities for other RRab's.  Although a moderately
large number of RRab's have known metallicity (eg. Layden 1994; Jurcsik \& Kovacs 1996), almost all the stars surveyed
are brighter than the SDSS $i=15$ threshold for spectroscopic observations. However, de Lee (2008) provides [Fe/H] for
more than 200 RRab's observed over a range of metallicities.  This work comes from extensive analysis of SDSS and CTIO
spectra as well as from photometry-based metallicities using the Fourier method of Jurcsik \& Kovacs (1996), and
period-amplitude method of Sandage (2004).  Although De Lee (2008) undertook their own analysis of SDSS spectra,
comparison of SDSS DR8 values to those derived from the same underlying data would likely contain a significant bias.
Therefore, we restrict our comparison to the remaining 253 De Lee~(2008) values. For these sources we find a total of
190 SDSS spectra with [Fe/H] values. Of these, 15 sources have metallicity based on CTIO spectra, 57 use the Fourier
method on SDSS lightcurves, 26 are based on Fourier analysis of De Lee (2008) photometry, and 13 use the photometric
analysis from the Sandage (2004) period-amplitude method.  In Figure \ref{Fedelee}, we plot a comparison between the
SDSS metallicities and those from De Lee (2008).  In addition, we include the 26 globular cluster RRab's with SDSS
metallicities.  Linear regression of the data gives:

\begin{equation}\label{FS}
[Fe/H] = 0.828 \times [Fe/H]_{SDSS} - 0.408.
\end{equation}

\noindent 
The overall result shows that the SDSS RRab metallicities are overall slightly higher 
than expected.
After subtracting the linear fit from the data the level of scatter matches that observed for repeated SDSS observations
($\sigma =0.22$).  This suggests that the SDSS values are characterized by this level of uncertainty. It may be possible
to obtain more accurate values of metallicity using values derived from individual SDSS exposures as noted by De Lee
(2008).  In this way one could correct for the phase of the SDSS observations. An alternate method for determining RRL
metallicities observed random phases has been developed by For et al.~(2011).  Reanalysis of the spectra using this
method may also yield improved results.

In Figure \ref{FE}, we present the distribution of RRL metallicities derived from SDSS spectra corrected by equation
\ref{FS}.  The distribution itself peaks near $\rm [Fe/H] = -1.55$ and exhibits a long tail extending to very low
metallicities.

\subsubsection{Radial Velocities}

As we noted above, there is well known dependence of radial velocity on observational phase.  The size of these variations
has been measured from repeated observations of RRLs (eg. For et al.~2011). Apart from small uncertainties in
heliocentric corrections, the difference between repeated observations of RRab's will reflect the pulsational variation.
Therefore, following our metallicity analysis we determined differences in the velocity measurements between pairs of
SDSS spectra. In Figure \ref{RadPul}, we present a histogram of these velocity variations.  Fitting a Gaussian to the
distribution we find $\sigma=25$ km/s. As expected this dispersion in much greater than the uncertainties quoted by 
SDSS for the individual radial velocity measurements.

As SDSS observations are composites from multiple exposures, the pulsational signal from an RRab can be washed out if
spectra are combined from varying phases.  To determine radial velocities of the RRab's we initially extracted the SDSS
observation start and end times from the SDSS DR8 database. We then removed SDSS spectra where the difference between
the start and end time of the observations was greater than three hours.  This left a set of 1239 spectra of the 
original 1871. In most cases the total time span of the remaining observations was around an hour.

Apart from problems of spectra being taken an indeterminate times one must consider the importance of the phase at which
the observations were taken.  Sesar et al.~(2012) note that observations taken after phase 0.95 exhibit rapid velocity
variations and therefore are have uncertain velocity corrections. Using our Fourier fits to the RRab lightcurves we
derive the phase range over which the RRab spectra were taken. The average phase length of SDSS observations for the
1239 remaining RRab spectra was 0.087.  Since there is uncertainty in the exact phases of the RRab's at time of the SDSS
observations (due to uncertainties in their period), we decided to remove RRab's with SDSS spectra that began before
phase 0.1, or ended after phase 0.95.  The final set consists of 905 spectra (less than half of the original set).

To accurately correct for velocity variation one must also consider how the radial velocities were measured.  SDSS
radial velocities are derived from both the metallic and hydrogen lines. Sesar et al.~(2012) recently noted differences
between velocities measured using hydrogen and metallic lines as references and derived relationships for correcting 
these. These authors found that the combination of three Balmer lines would lead to uncertainties of a few km/s.  We compared
corrections based on combinations of Balmer as well as metallic lines and found typical differences of less than 10km/s.
We combined the relationships given by Sesar et al.~(2012) to produce an appropriate correction for the SDSS measurements.
Using the Sesar et al.~(2012) velocity curves and amplitudes we then determined pulsation corrections for each of the 
905 spectra by averaging the velocities over the period between the start and end phase of the SDSS observations.

After correcting for pulsation velocities we redetermined the distribution of velocities for objects that had 
repeated observations. Because there are fewer spectra in the reduced set the number of repeat observations
is greatly diminished. In Figure \ref{RadPulcor}, we plot the resulting distribution. A Gaussian fit to the distribution
gives $\sigma= 14.3km/s$. Clearly the pulsational velocity corrections have improved the agreement between successive
measurements of the radial velocities. We adopt this level of uncertainty for all the remaining spectra and 
do consider the other spectra. We also recalculated variations in the metallicities for these 905 spectra and 
found no change in the standard deviation.

Analysis of radial velocities based on individual SDSS exposures has been performed by De Lee (2008) for
RRab's in found in stripe 82. Application of this technique can remove problems associated with composite 
SDSS spectra. Indeed, additional analysis of SDSS RRab spectra is underway by De Lee et al.~(2012, in prep). 
This work should be able to recover radial velocities for the RRab's excluded here.

In order to understand the RRab radial velocities in the context of Galactic structure, we follow Law \& Majewski (2010,
hereafter LM10) and transform the velocities to the Galactic standard of rest (GSR). We assume a Solar peculiar motion
of (U, V, W) = (9, 12 + 220, 7) km/s in the Galactic Cartesian coordinate system.  In Figure \ref{Vel}, we plot the
distribution of velocities using a 20 km/s bin size. A Gaussian fit to the distribution gives mean $\rm \bar{V_{GSR}} =
-18.3\,km/s$ and dispersion $\rm \sigma =119.0\,km/s$.  The distribution appears to show some non-Gaussian structure.
However, there is likely some observational bias caused by preferentially detecting nearby sources.

\section{RRab Distances}

The absolute magnitudes of RRab are given by Catelan \& Cort\'es (2008):

\begin{equation}
M_V = 0.23 \times {\rm [Fe/H]}_{\rm ZW84}+ 0.948,
\end{equation}

\noindent where ${\rm [Fe/H]}_{\rm ZW84}$ is the metallicity in the Zinn \& West (1984) scale.
The average metalicity for our RRL with SDSS matches is $\rm [Fe/H]= -1.48$.  Thus we adopt an average magnitude
$M_V=0.61$.  This value is close to the value of 0.6 adopted by Keller et al.~(2008) and Sesar et al.~(2010). Like the
SDSS photometry the CSS V magnitudes were corrected for extinction using Schlegel et al.~(1998) reddening maps. The
dispersion in the metallicity is approximately 0.3 dex, which corresponds to a variation of 0.07 magnitudes. 
The uncertainties in RRab absolute magnitudes are sometimes noted as $\sim 0.05$ mags. However, the level 
of agreement between independent measurements (eg. Benedict et al. 2011, table 10) suggests true uncertainties 
are closer to $\sim 0.1$ mags. The distances to individual sources are determined using:

\begin{equation}
d = 10^{((V_0)_{\rm s} - M_V + 5)/5} 
\end{equation}

Here we have corrected the average RRab $V_0$ magnitudes to static values $(V_0)_{\rm s}$ using 
values derived from a polynomial fit to the amplitude corrections given by Bono et al.~(1995).
Combining the uncertainties from the photometric calibration and colour variation of 0.09 with the
variations in metallicity and uncertainty in RRab absolute magnitudes, we derive an overall 
uncertainties of 0.15 magnitudes. This corresponds to a $\sim 7\%$ uncertainties in distances. 
From our photometric calibration it is clear that faint RRL will have larger uncertainties 
in average magnitude. However, these uncertainties should generally not exceed 0.25 magnitudes 
($\sim 12\% $ in distance). 
The faintest CSS RRab's in our dataset have $V \sim 19.5$, corresponding to 60 kpc.
In Table 2, we present the u,g,r,i,z magnitudes, metallicities and radial velocities
transformed to the Galactic standard of rest for sources covered by SDSS DR8.

\section{Galactic Structure}

As the Sgr stream is near the ecliptic plane and our spatial coverage is complete in ecliptic longitude ($\lambda$) (apart
from the Galactic plane region), in Figure \ref{Eclip} we plot the distribution of RRab distances and magnitudes versus
$\lambda$.  The Sgr streams are relatively clear in both plots. However, the magnitude plot gives a clearer picture
of the inner (trailing) stream, showing two clear arms of the Sagittarius remnant at relatively small Galactocentric distances.
One arm extends from $V \sim 19.25$ (54 kpc) at $\lambda = 225\arcdeg$, to $V \sim 17$ (19 kpc) at $\lambda =
120\arcdeg$.  The other goes from $V \sim 18.5$ ($d=38$kpc) at $\lambda \sim 60\arcdeg$, to $V\sim 17$ ($d=19$ kpc) at
$\lambda = 305\arcdeg$.

Of the RRL, 11019 are at Galactocentric distances $d_G < 33.5\, {\rm kpc}$ and 1208 are beyond that. Since our detection
completeness for the distant RRab's is much lower than the nearby brighter sample it is clear that there are a 
significant number of RRL within the Galactic halo.

In Figure \ref{FeSgr}, we plot the distribution of spectroscopic metallicities for the 219 RRab's with $d_{G} > 33.5$
kpc. Each value has been corrected via equation \ref{FS}. The distribution appears slightly more metal-poor than the 
overall distribution. The dispersion remains the same at around 0.3 dex and suggest the objects are a mixture rather 
than a single population. A number of sources near $\rm [Fe/H] = -2.2$.

These halo RRab have significantly lower metallicities than observed for M-giants in the Sgr stream given by LM10 ($\rm
\langle[Fe/H]\rangle \sim -0.9$). However, M-giants and RRL's a know to trace different populations. Nevertheless, Casey et al.
(2012) discovered a number of Sgr K-giants with $\rm \langle[Fe/H]\rangle \sim -1.7$. With significant the level of
uncertain the result is also consistent with Stripe-82 RR Lyrae in the Sgr stream measured by Watkins et al.~(2009) 
with $[Fe/H]=-1.41\pm 0.19$ and that of $\rm [Fe/H]=-1.76 \pm 0.22$ measured by Vivas, Zinn \& Gallart (2005) for 
12 Sgr stream RRL.

\subsection{Comparison with the LM10 Sagittarius Model}

In order to compare our results further with models of the Sgr stream we first select the 905 RRab's with SDSS radial
velocity uncertainties of $\sigma= 14.3km/s$. We then find sources with Galactocentric distances $> 30\, {\rm kpc}$ and
coordinates within range of Sagittarius stream based on the Majewski et al.~(2003) Sgr stream coordinates ($-15 < B <
-15\arcdeg$).  As a comparison, we select LM10 data points in this same region.  In Figure \ref{VelSgr}, we compare the
SDSS data with this LM10 model. The bulk of sources are a very good match to the model. However, compared to the model,
the RRab's in the region $110 < \alpha < 180 \arcdeg$ appear to show a velocity trend with much large velocities 
than the LM10. The region $250 < \alpha < 360 \arcdeg$ is not plotted since it is very poorly covered by SDSS data.

To compare the distant RRab distribution with the LM10 model we determine Galactocentric distances for all
the objects. We next separate the RRab's into four groups, namely a sample at distances $r_G < 33.5$ kpc and three more
distant halo RRab samples with ranges 33.5 to 38 kpc, 38 to 45 kpc, and 45 to 65 kpc. These distance ranges were chosen
to broadly separate the 1468 RRab's at distances $> 33.5$ kpc into three similar groups of $\sim 500$ RRL each. In Figure
\ref{CompLaw}, we plot the distribution of the Halo RRab's among the sample along with the LM10 model.  The
Sgr stream is clearly visible. However, many RRab's are seen in the region $110 < \alpha < 180$ beyond 45 kpc and
are not explained by the LM10 model. Additionally, many distant RRL are found in locations not expected from the
Sgr model. There is no obvious division of the Sgr RRab's into two streams as discovered by Belokurov et al. (2006)
and Koposov et al.~(2012).

To further investigate the Sgr stream, based on Figure \ref{Eclip}, we select the clear Sgr RRab with $\rm d_h > 30 kpc$
in the region of $-41\arcdeg < B < 31\arcdeg$. We determined the density distribution and plot this in Figure \ref{DistSgr}.  After
binning the data in two degree bins we find that the main density distribution is well described by a single Gaussian centered
at $\rm B=-1.4 \pm 0.3\arcdeg$, with $\sigma = 6.8 \pm 0.3 \arcdeg$, plus a background of $13.8 \pm 1.6$ RRab's.  Here
the number of Halo RRab's is a factor of $\sim 200$ smaller than main-sequence turn off (MSTO) stars analyzed by Koposov
et al.~(2012). Thus, although there is no obvious evidence for a second peak near $B= -8 \arcdeg$, we can
conclude that the RRab are distributed across the two streams (seen in SDSS MSTO stars and 2MASS M-giants),
rather than limited to one.

In Figure \ref{DistRA}, we plot the distribution of heliocentric distances for the RRab's.  The Sgr streams are clearly 
seen rising up to heliocentric distances of $\sim 52$ kpc near $\alpha = 230 \arcdeg $ and 30kpc near $\alpha = 70 \arcdeg$. 
We also include the M-giant selected by LM10 as leading and trailing Sgr stream sources. The RRab's appear to have 
distances consistent with the M-giants, although with significantly less scatter. We note that Newberg et al.~(2003) 
found a 13\% difference in the distance to the Sgr stream when comparing A-coloured stars, such as RR Lyrae and 
BHB stars, to M-giants, which is not confirmed in this study.

In Figure \ref{DistRA}, we also compare the results with the Sgr stream N-body model of LM10. For an improved comparison
we apply extinction to each of the LM10 data points and remove sources that would have apparent magnitudes corresponding
to RRab's beyond our detection limit ($d_h \sim 60$ kpc).  We also remove points within $15 \arcdeg$ of the Galactic
plane, since this region is not covered by CSS data.  Although there is some overall agreement, at distances less than
20 kpc the presence or Galactic halo RRL is a significant factor, making comparison with the predicted nearby streams
difficult. This is particularly the case near the Galactic center. We over-plot two lines that clearly demonstrate the
difference in distances between
the LM10 N-body model and the observational data. 
The model predicts stars $\sim 5$ kpc further than observed. This is expected since LM10 
found that their model predicted fainter $K_s$ magnitudes for M-giants than observed.

Other differences include the Sagittarius leading arm near $d_h = 45$ kpc $\alpha = 235 \arcdeg$. 
Although there appear to be stars near this location, the density is much lower than predict by 
the LM10 model. 

The distances of RRab in Sgr leading arm are found to vary by up to $\sim 10 kpc$ (corresponding to 0.37 mags). As
$3\sigma$ uncertainties in the RRab magnitudes are $\sim 0.3$ mags, the intrinsic depth of the leading Sgr stream 
is expected to be significantly less than 10 kpc.

Furthermore, there is only a weak sign for a Sgr leading stream to the dense region at $RA=110 \arcdeg$, $d_h = 45$ kpc.
There is some evidence for RRab at a greater distance.  However, we note that in this region of the Sagittarius stream
is not very well constrained by our observations.  As expected we do observe the trailing stream in the region $0 <
\alpha < 80 \arcdeg$. Without any doubt this stream does not continue on the Sgr stream in the region $110 < \alpha <
240\arcdeg$.

\section{Discussion and Conclusions}

We have presented the initial results of a survey of public CSS data (CSDR1) for RRab's and discovered $> 12000$,
of which $\sim 9400$ are newly discovered. The full sample of RRab's range in brightness $12.5 < V < 19.5$ and thus reaches 
heliocentric distances from 3 to 60 kpc. The objects in this catalog generally have average absolute $V$
magnitudes with uncertainties $< 0.1$ mag and periods accurate to better than $0.01\%$. More than half of the sources
have accurate five-colour photometry from the SDSS DR8 release and 1531 have SDSS spectra.

Although this set of RRab's is incomplete within the survey area (because of the magnitude, sampling and reddening
limits of the data), this data set provides a source for more detailed studies of halo streams and structures, as well 
as a means of constraining the shape, mass and extent of the Galactic halo (Koposov et al. 2012, in prep).

Here we provide a first comparison with models of the Sagittarius stream, a find a few discrepancies that should
eventually lead to a better understanding of the formation history of the Milky Way halo and its dwarf satellite
galaxies. The current data set reaches declinations $-22\arcdeg$ which extends it $sim 20\arcdeg$ beyond the limits 
of most sources in the SDSS survey. 
More than 10,000 RRab have been discovered in photometry from SSS survey in the region $-75\arcdeg <
\delta < -22 \arcdeg$ (Torrealba et al. 2012, in prep). The combination of CSS and SSS RRab will provide probes 
of halo structure and Galactic potential covering $\sim 75\%$ of the sky. Additionally, although MLS survey 
covers much less area than CSS and SSS, it probes the halo RRL to distances beyond 100 kpc and has recently 
confirmed the presence of a distance tidal stream overlapping the Sgr system (Drake et al.~2012, in prep.).

\acknowledgements 
CRTS and CSDR1 are supported by the U.S.~National Science Foundation under grants AST-0909182 and CNS-0540369.
The work at Caltech was supported in part by the NASA Fermi grant 08-FERMI08-0025, and by the Ajax Foundation. The CSS
survey is funded by the National Aeronautics and Space Administration under Grant No. NNG05GF22G issued through the
Science Mission Directorate Near-Earth Objects Observations Program. J. L. P. acknowledges support from NASA through
Hubble Fellowship Grant HF-51261.01-A awarded by the STScI, which is operated by AURA, Inc.  for NASA, under contract
NAS 5-26555.  
Support for M.C. and G.T. is provided by the Ministry for the Economy, Development, and Tourism's Programa Inicativa
Cient\'{i}fica Milenio through grant P07-021-F, awarded to The Milky Way Millennium Nucleus; by Proyecto Basal
PFB-06/2007; by Proyecto FONDECYT Regular \#1110326; and by Proyecto Anillo ACT-86.
SDSS-III is managed by the Astrophysical Research Consortium for the Participating Institutions of the SDSS-III
Collaboration Funding for SDSS-III has been provided by the Alfred P. Sloan Foundation, the Participating Institutions,
the National Science Foundation, and the U.S. Department of Energy Office of Science. The SDSS-III web site is
http://www.sdss3.org/.  This research has made use of the International Variable Star Index (VSX) database, operated at
AAVSO, Cambridge, Massachusetts, USA



\begin{figure}{
\hspace*{-1.5cm}\epsscale{1.4}
\plotone{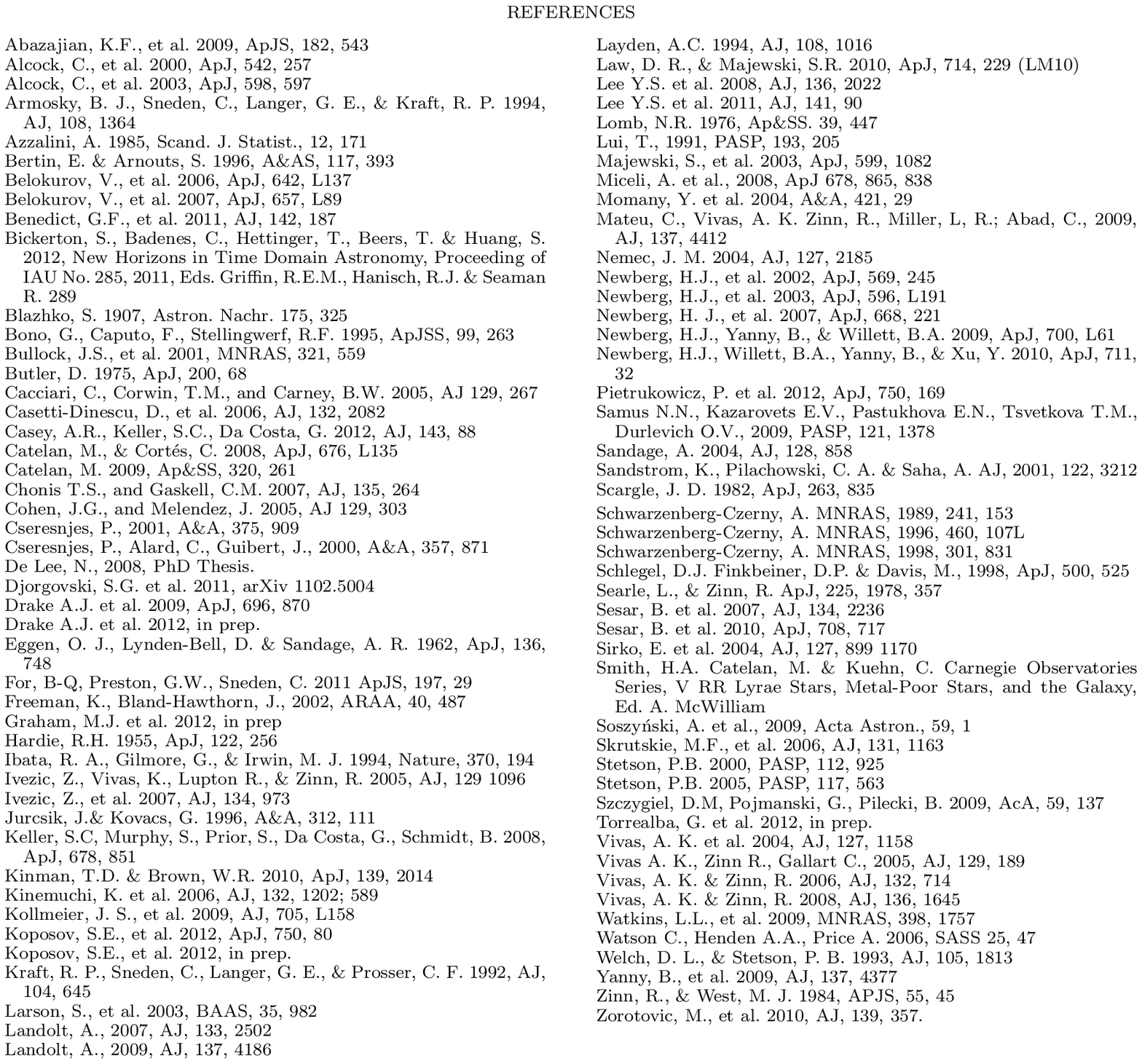}
}
\end{figure}

\newpage

\begin{figure}{
\epsscale{1.2}
\plotone{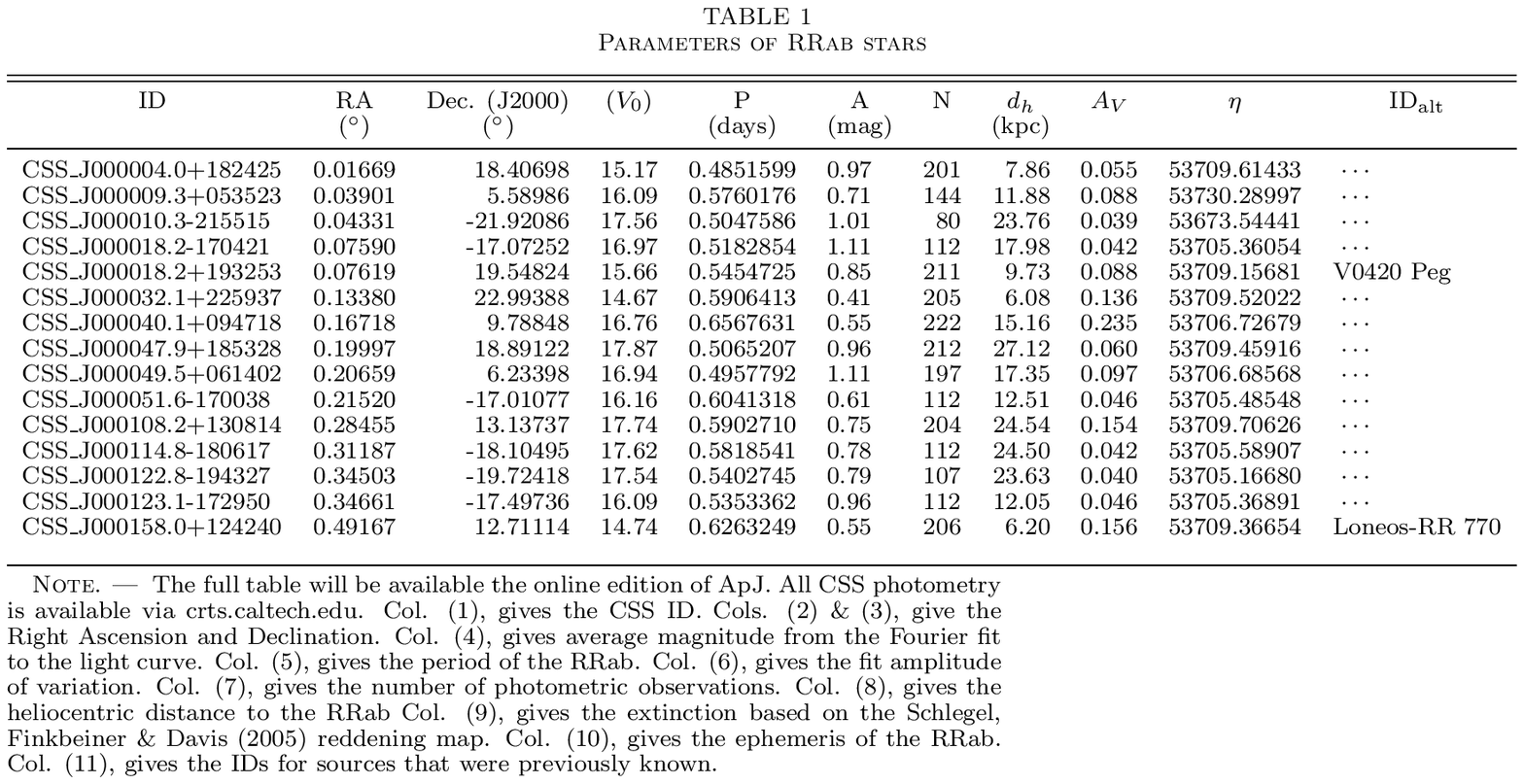}
}
\end{figure}

\begin{figure}{
\epsscale{1.0}
\plotone{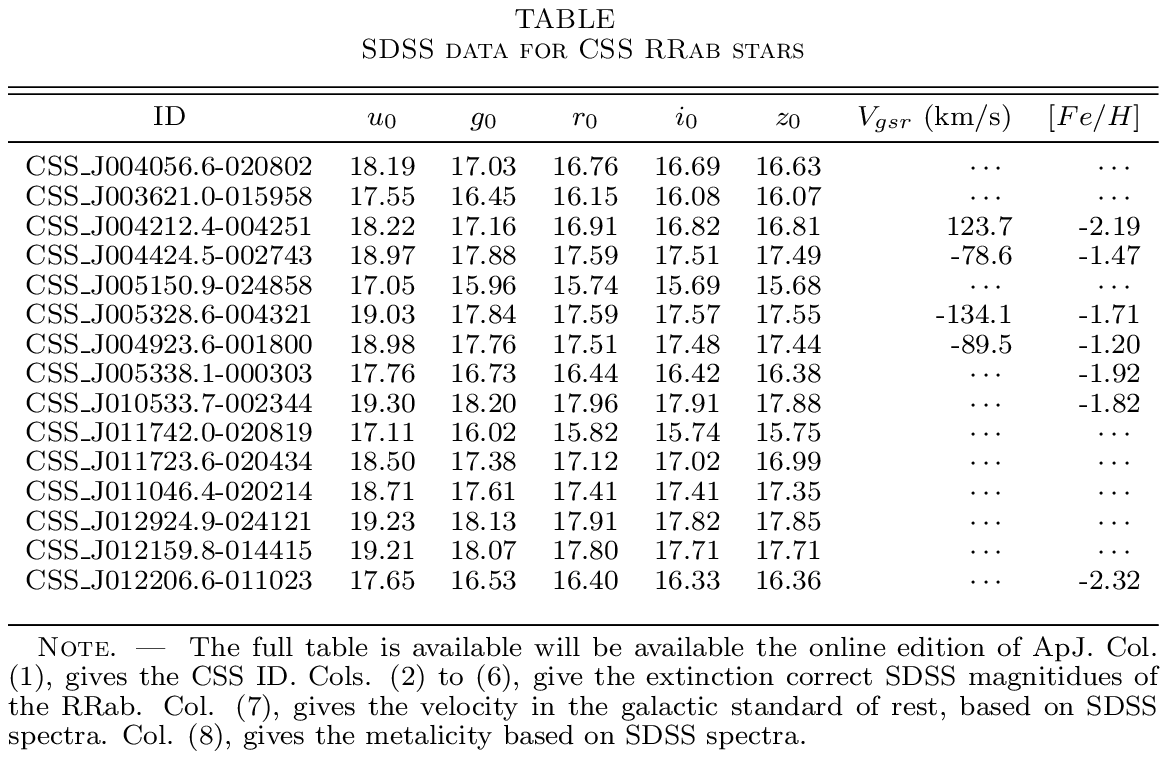}
}
\end{figure}
\clearpage

\begin{figure}{
\epsscale{0.6}
\plotone{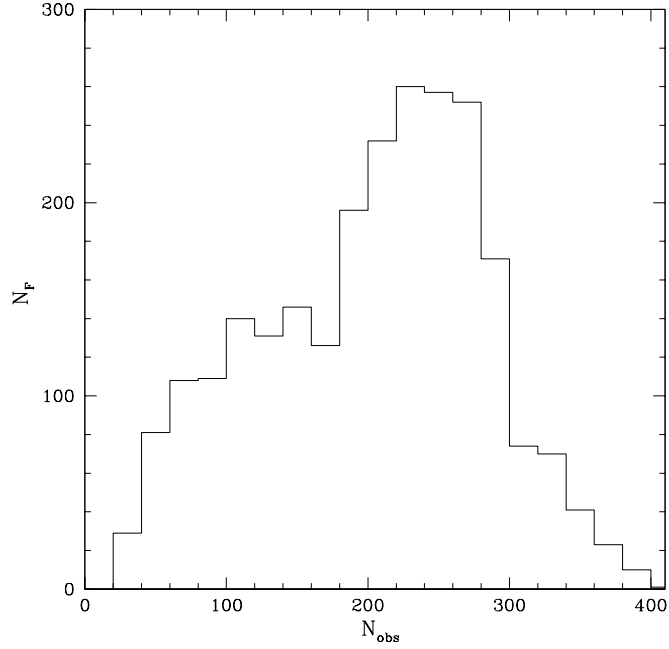}
\caption{\label{NfNo}
The distribution of the number of CSS image fields (N$_F$)
having specified numbers of observations (N$_{obs}$).
}
}
\end{figure}

\begin{figure}{
\epsscale{1.0}
\plottwo{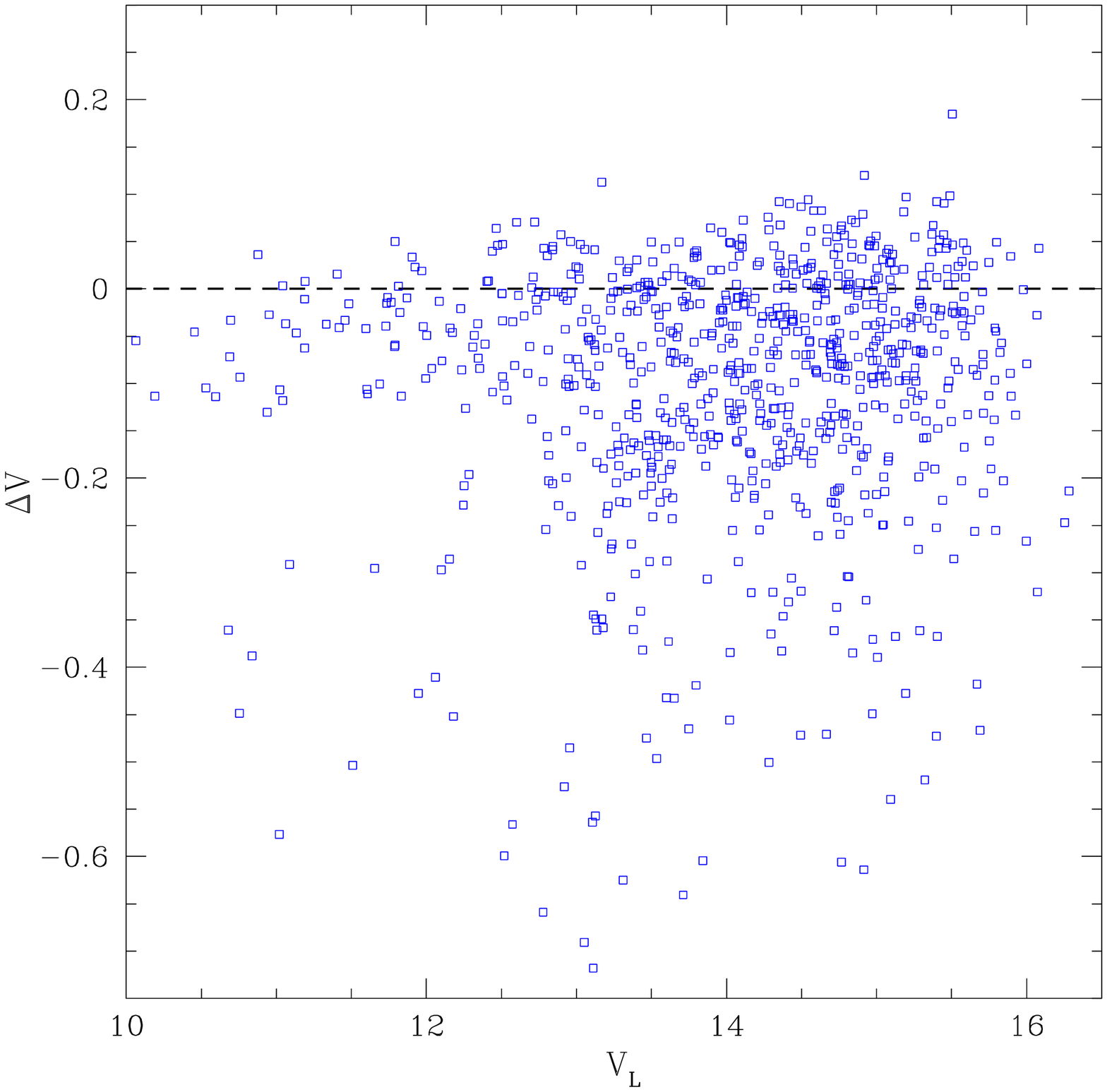}{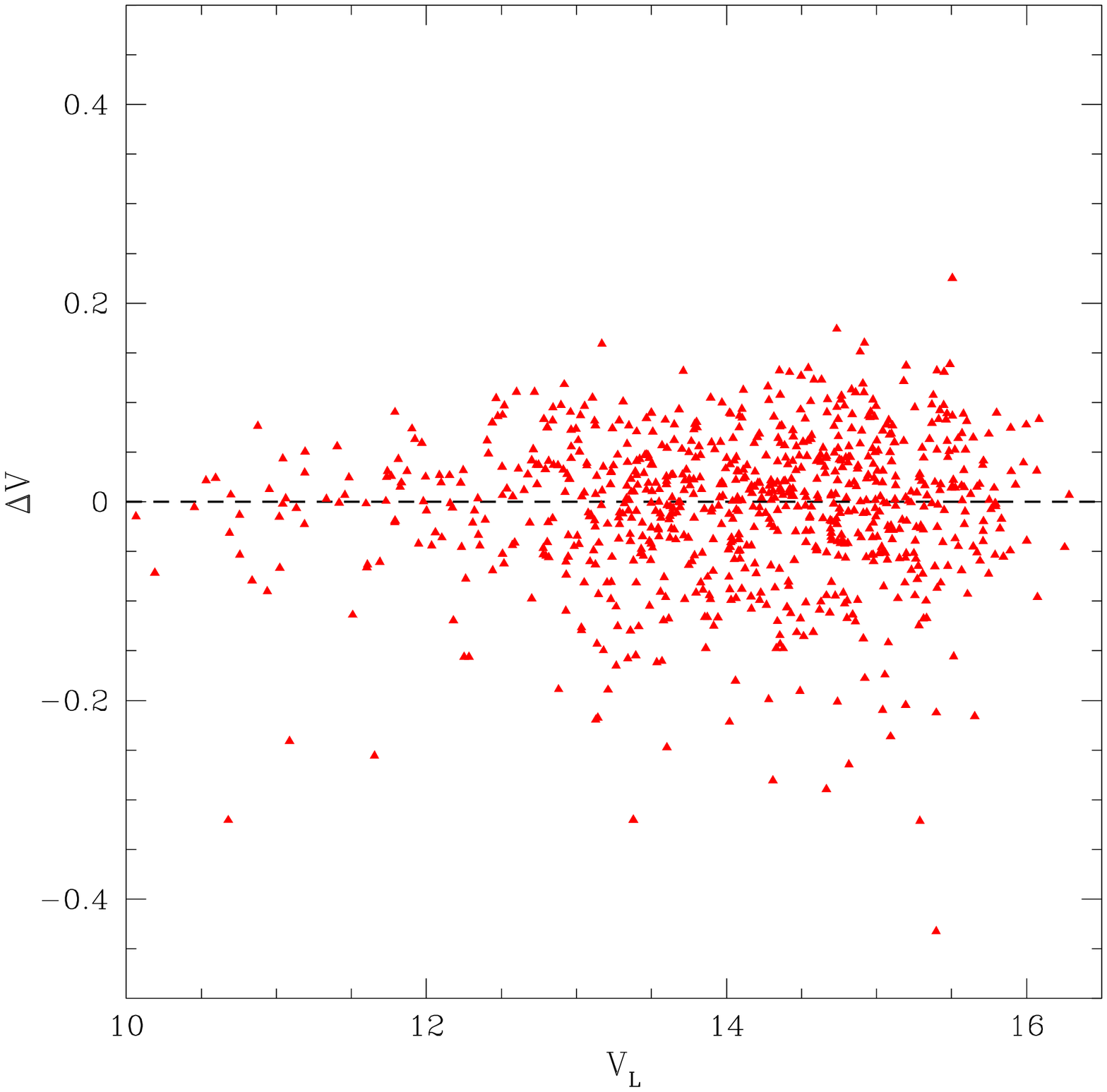}
\caption{\label{Landolt}
A comparison of CSS $V$ magnitudes with Landolt magnitudes for standard stars.
The left plot shows the difference in magnitudes before colour corrections
and the right plot shows the difference after colour corrections have been applied.
}
}
\end{figure}

\begin{figure}{
\epsscale{0.6}
\plotone{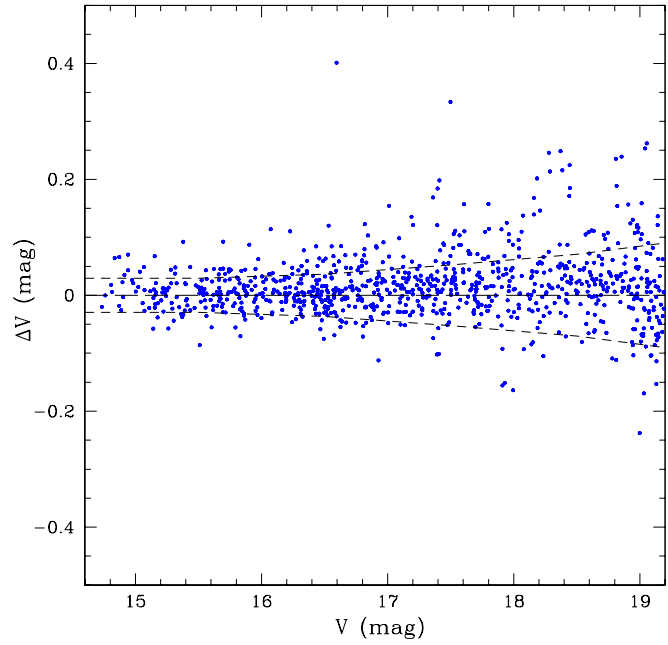}
\caption{\label{BHB}
A comparison between SDSS and CSS magnitudes transformed to $V$ for $\sim 1000$ 
BHB stars selected by Sirko et al.~(2004). The long dashed-line marks the expected 
zero-offset line. The short-dashed lines show the one $\sigma$ uncertainties.
}
}
\end{figure}

\begin{figure}{
\epsscale{0.6}
\plotone{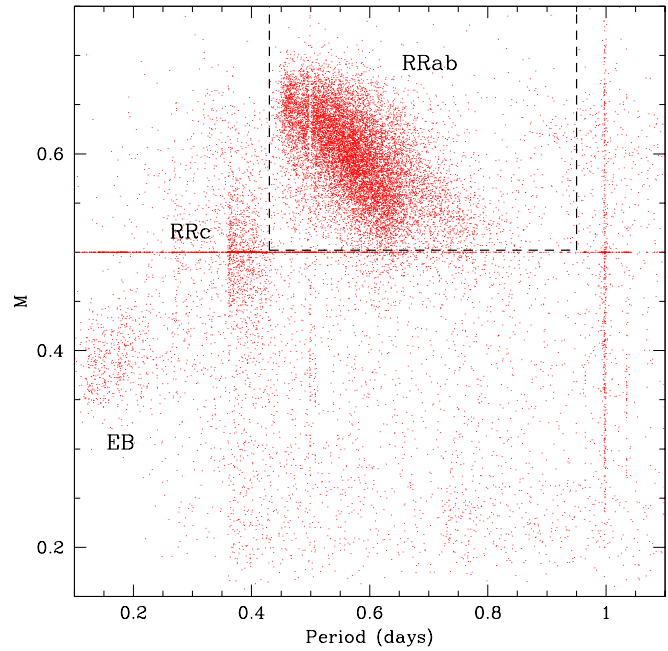}
\caption{\label{mtest} 
Values of the Kinemuchi et al.~(2006) M-test statistic used to select 
RRab's from other period variables. The dashed-line shows the border 
of the period region selected in this work.
}
}
\end{figure}

\begin{figure}{
\epsscale{1.0}
\plottwo{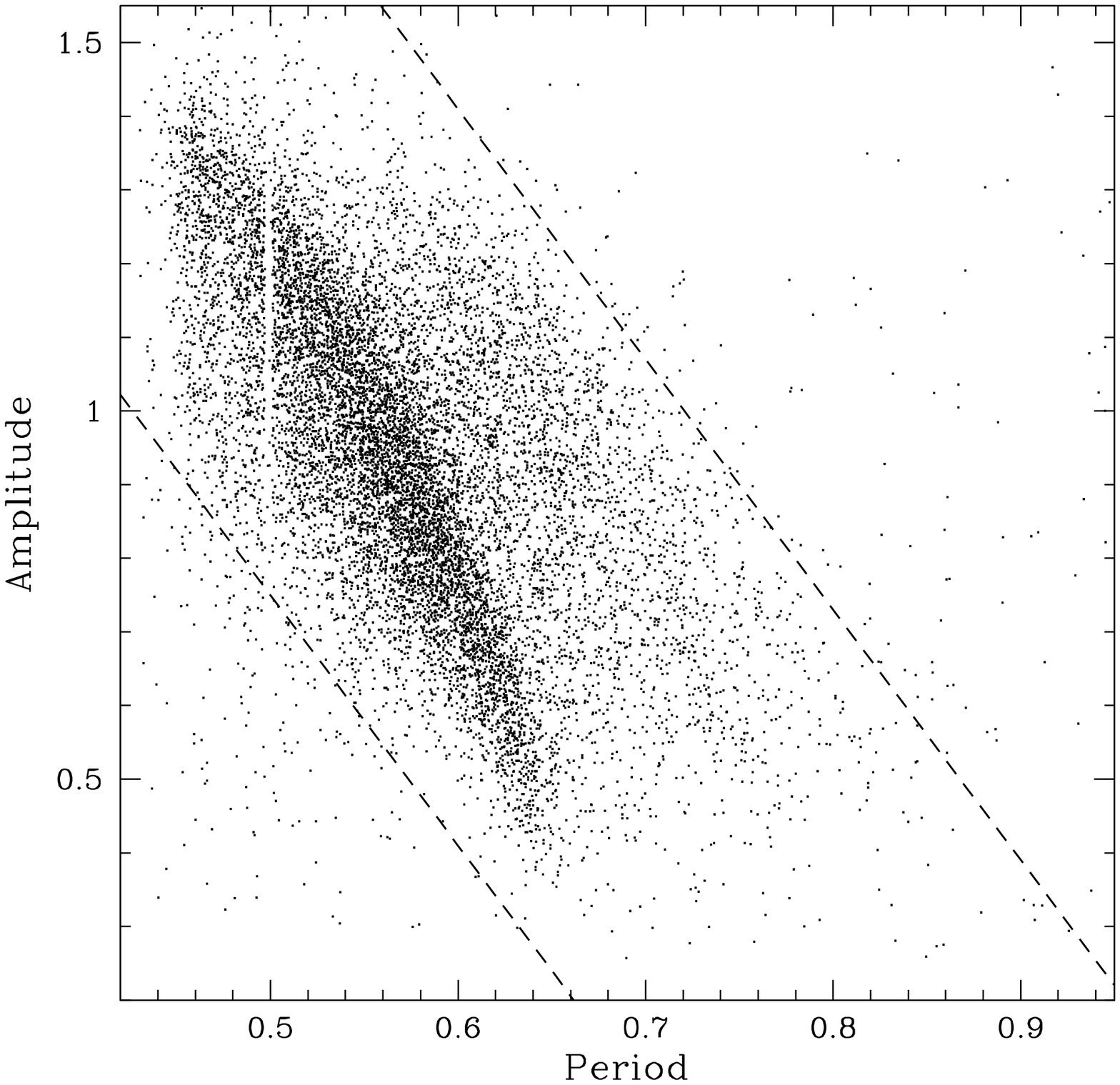}{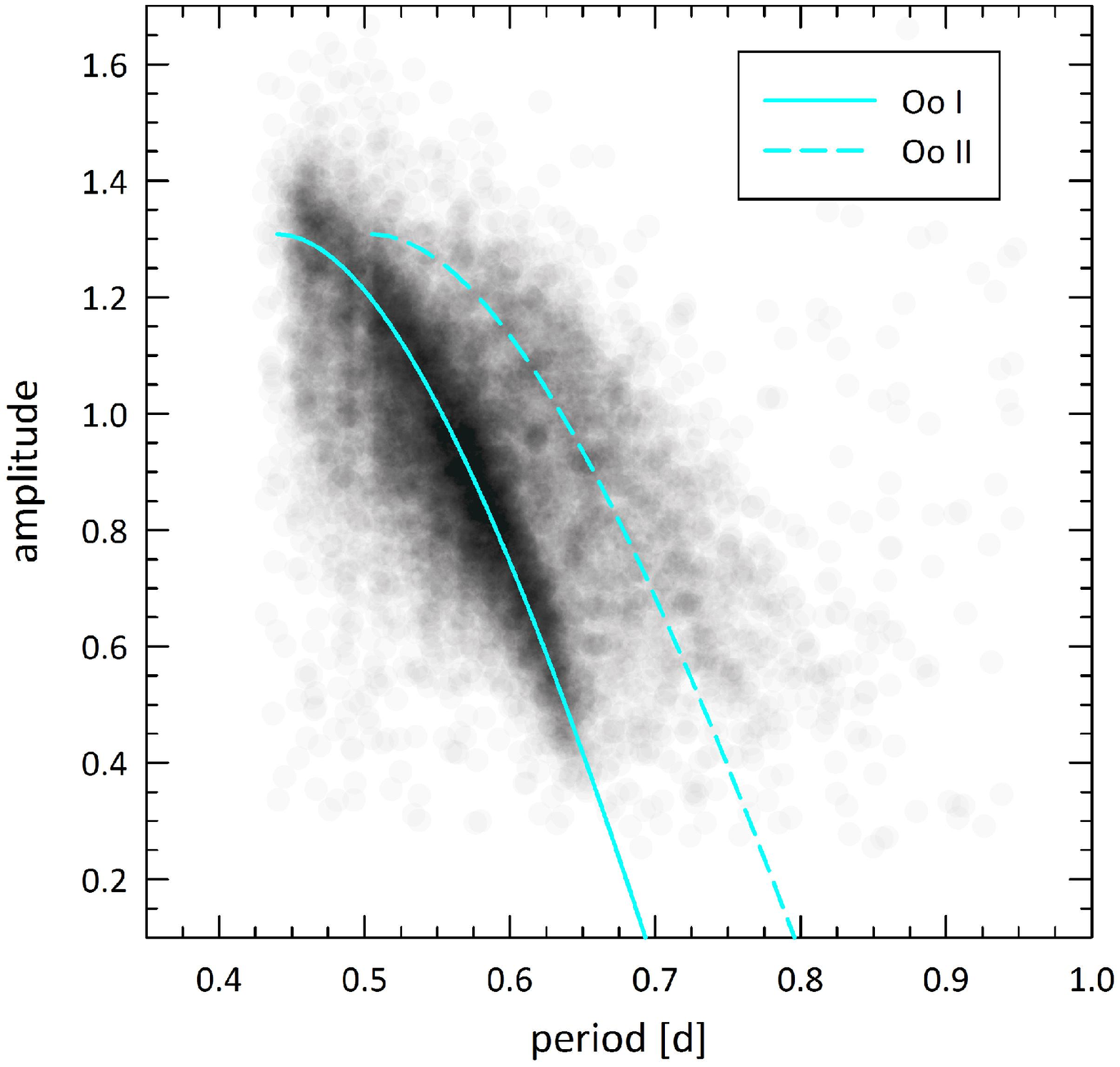}
\caption{\label{PerAmp} 
The Period-Amplitude diagram (aka ``Bailey diagram'') for RRab's found in this work.
In the left panel the dashed-lines presents the lower ($A > 2.3 -3.4 \times P$)  
and upper ($A = 3.3 - 3.4 \times P$) limits expected for most of 
the RRab's. The RRab's near 0.5 days are missing since the selection 
procedure removes non-periodic sources occurring sampling aliases.
In the right panel we plot the Hess (point density) diagram with
reference lines for OoI and OoII systems, based on eq. (11) in 
Zorotovic et al.~(2010).
}
}
\end{figure}

\begin{figure}{
\epsscale{0.7}
\plotone{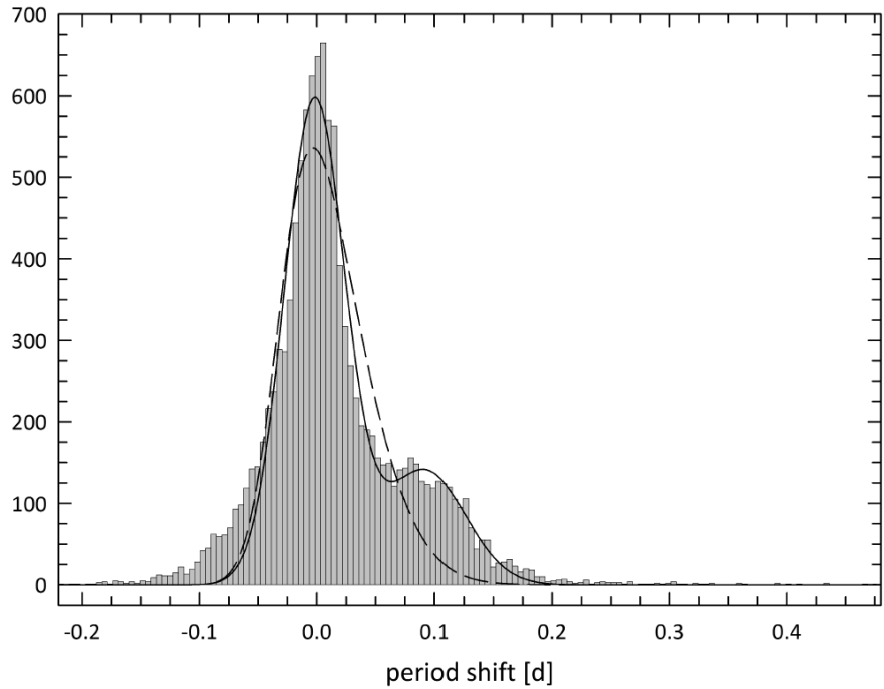}
\caption{\label{PerAmpM} 
The period-shift distributions for the CSS RRab's.
Here we plot the difference between the observed period and OoI 
period-amplitude line along with a two-Gaussian fit and a skew-normal 
fit to the resulting data.
}
}
\end{figure}

\begin{figure}{
\epsscale{1.0}
\plotone{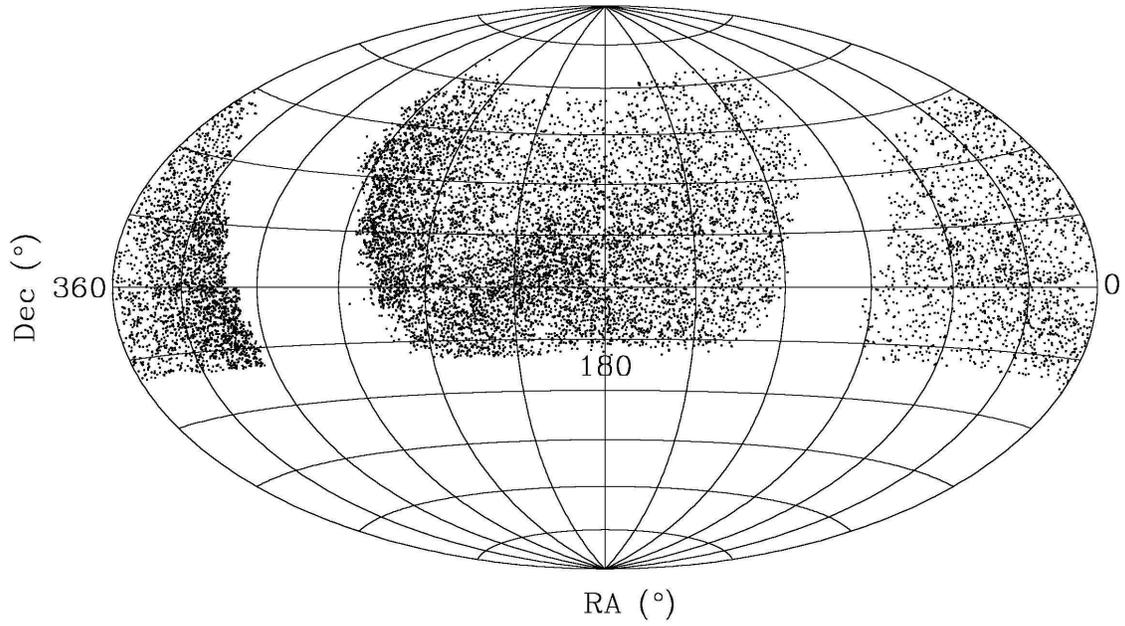}
\caption{\label{AitEq}
An Aitoff projection of the equatorial coordinates for all the CSS RRab
detected in this work. A higher resolution figure is available in the online
journal. 
}
}
\end{figure}

\begin{figure}{
\epsscale{1.0}
\plotone{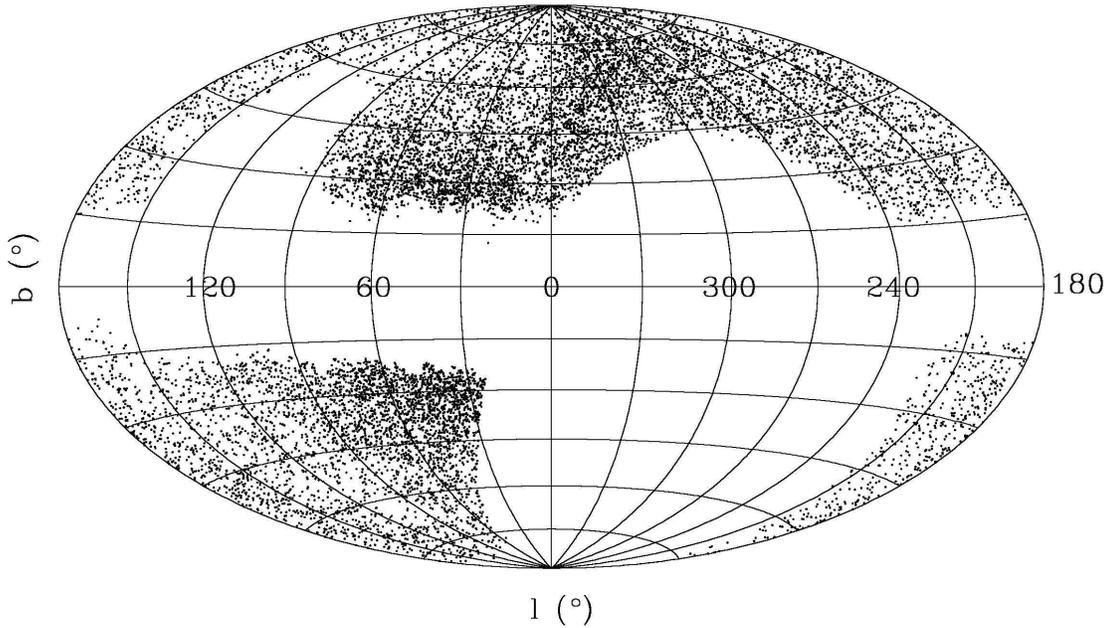}
\caption{\label{AitEqlb}
An Aitoff plot of the Galactic coordinates of all the CSS RRab 
detected in this work. The Galactic plane region with $|l| \lsim 15\arcdeg$
is avoided by CSS because of source crowding. A higher resolution figure is 
available in the online journal. 
}
}
\end{figure}

\begin{figure}{
\epsscale{1.0}
\plottwo{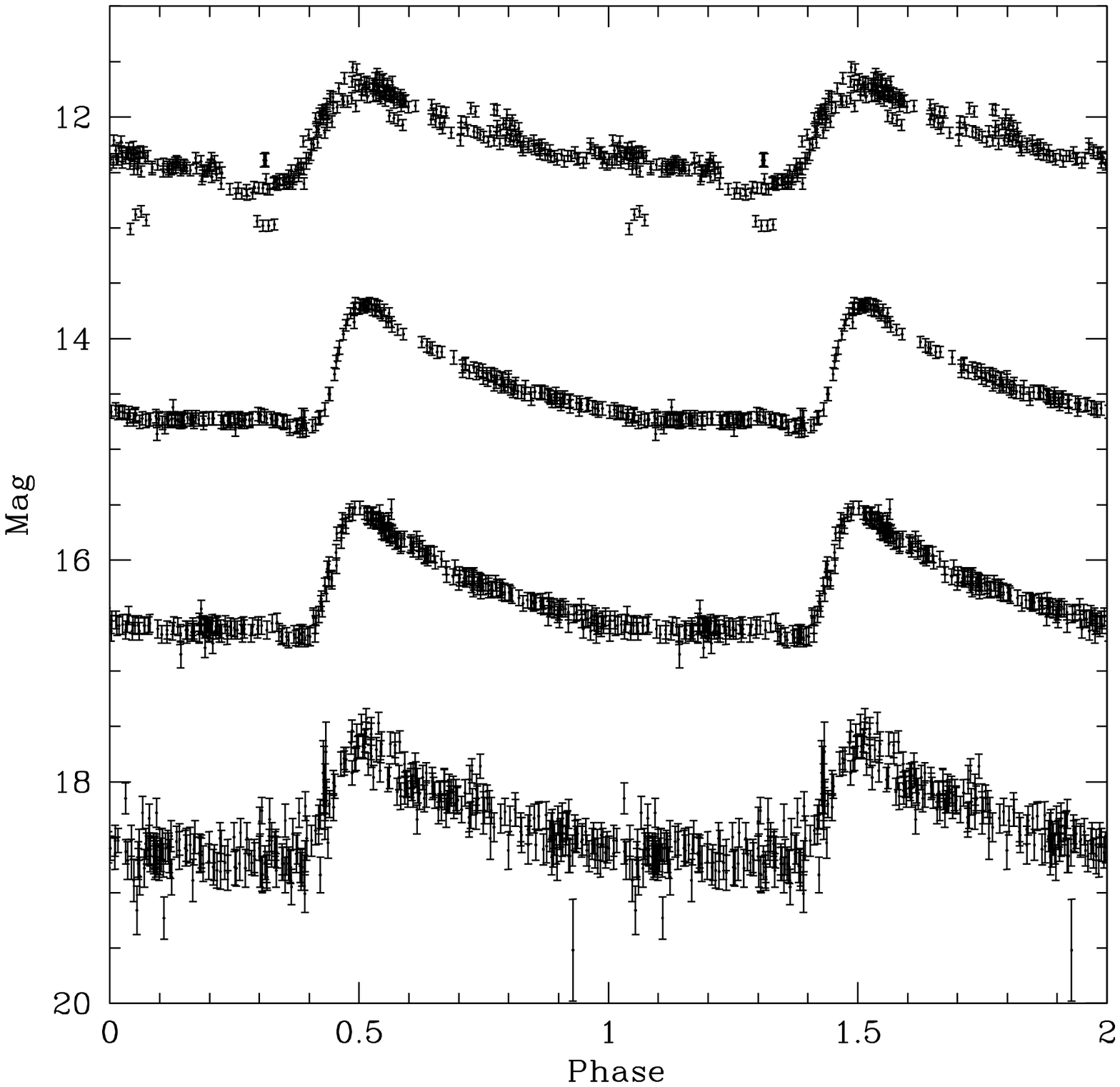}{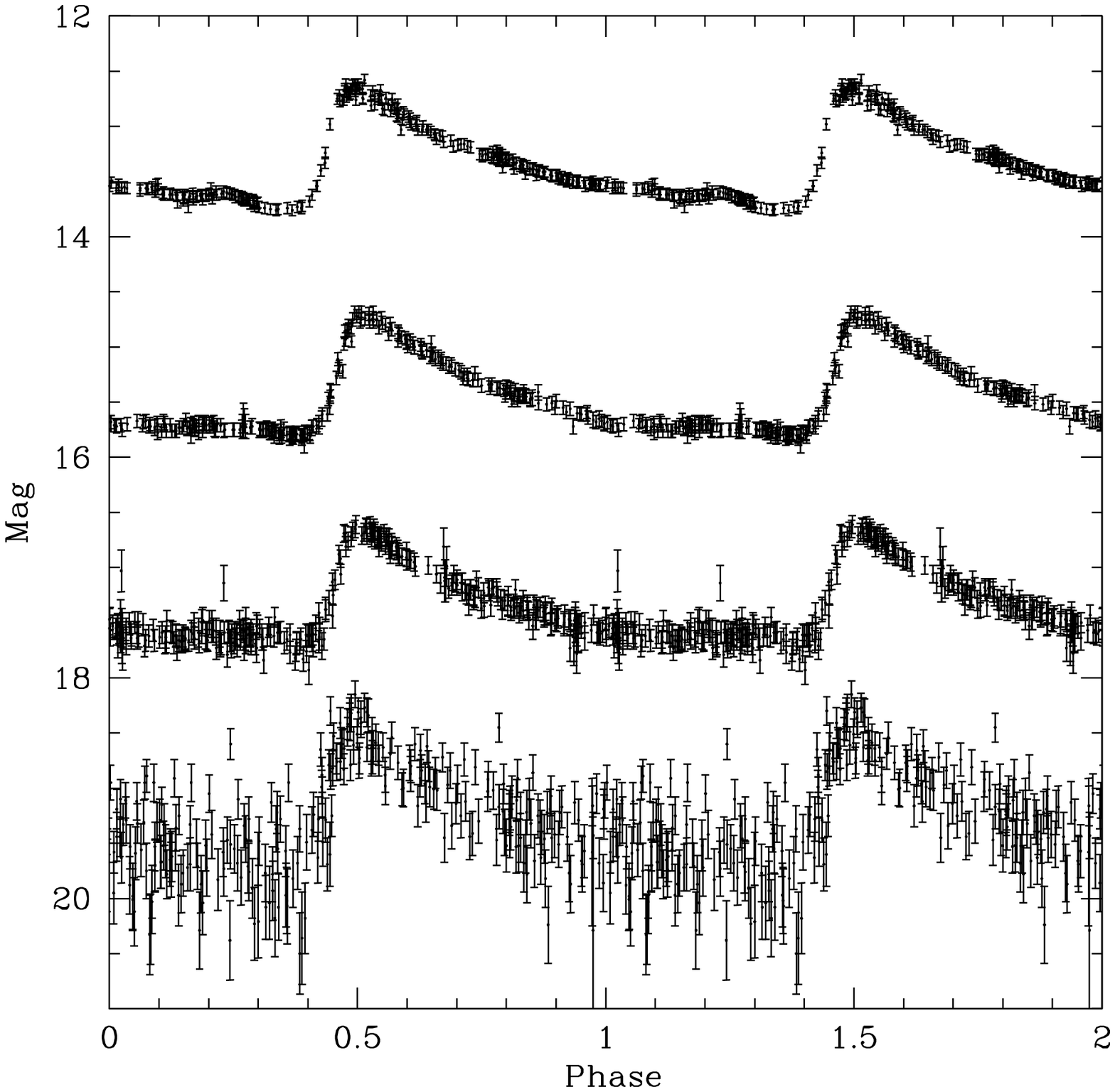}
\caption{\label{RRLC}
Examples of CSS RRL lightcurves phase folded with their best-fit 
periods. In the left panel we plot the lightcurves of RRab's
with average magnitudes of $V \sim$ 12.5, 14.5, 16.5 and
18.5. In the right panel we plot RRab's with magnitudes
of $V \sim$ 13.5, 15.5, 17.5 and 19.5.
}
}
\end{figure}

\begin{figure}{
\epsscale{0.6}
\plotone{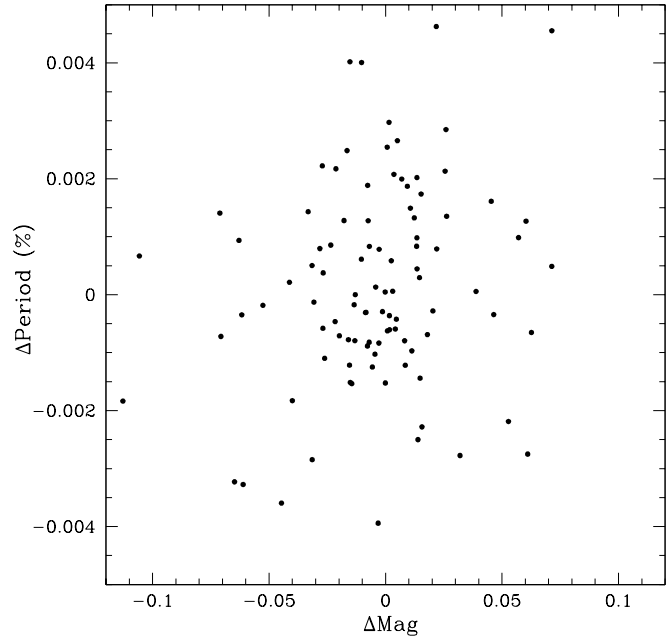}
\caption{\label{Diff}
The distribution of the differences in the period and average magnitude $V$
for 100 CSS RRab's that were detected in overlapping CSS fields.
}
}
\end{figure}
\clearpage

\begin{figure}{
\epsscale{0.6}
\plotone{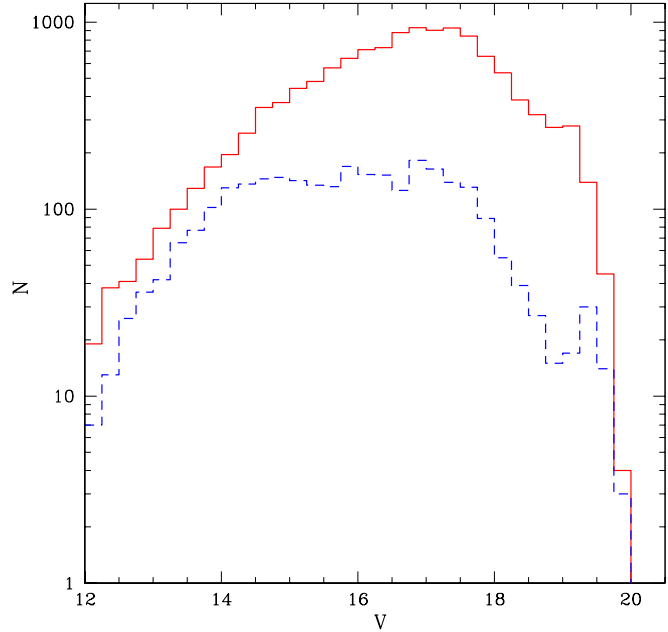}
\caption{\label{Hist} 
The magnitude distribution of CSS RRab's.
The solid line shows a histogram of the average $V$ magnitudes for RRab's discovered in CSS data.
The dashed-line shows the distribution of previously known RRab's recovered in this work.
}
}
\end{figure}

\begin{figure}{
\epsscale{1.0}
\plottwo{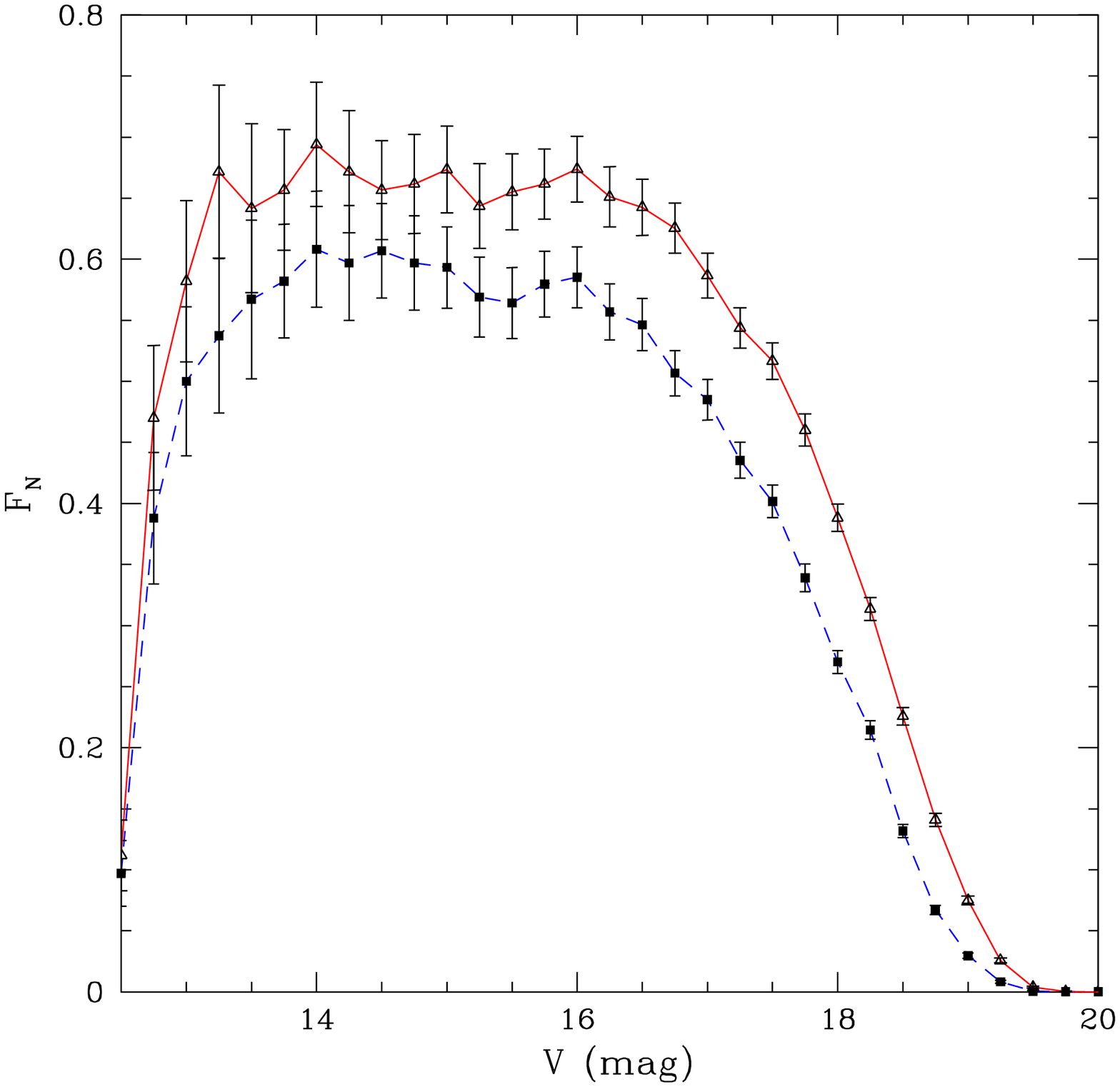}{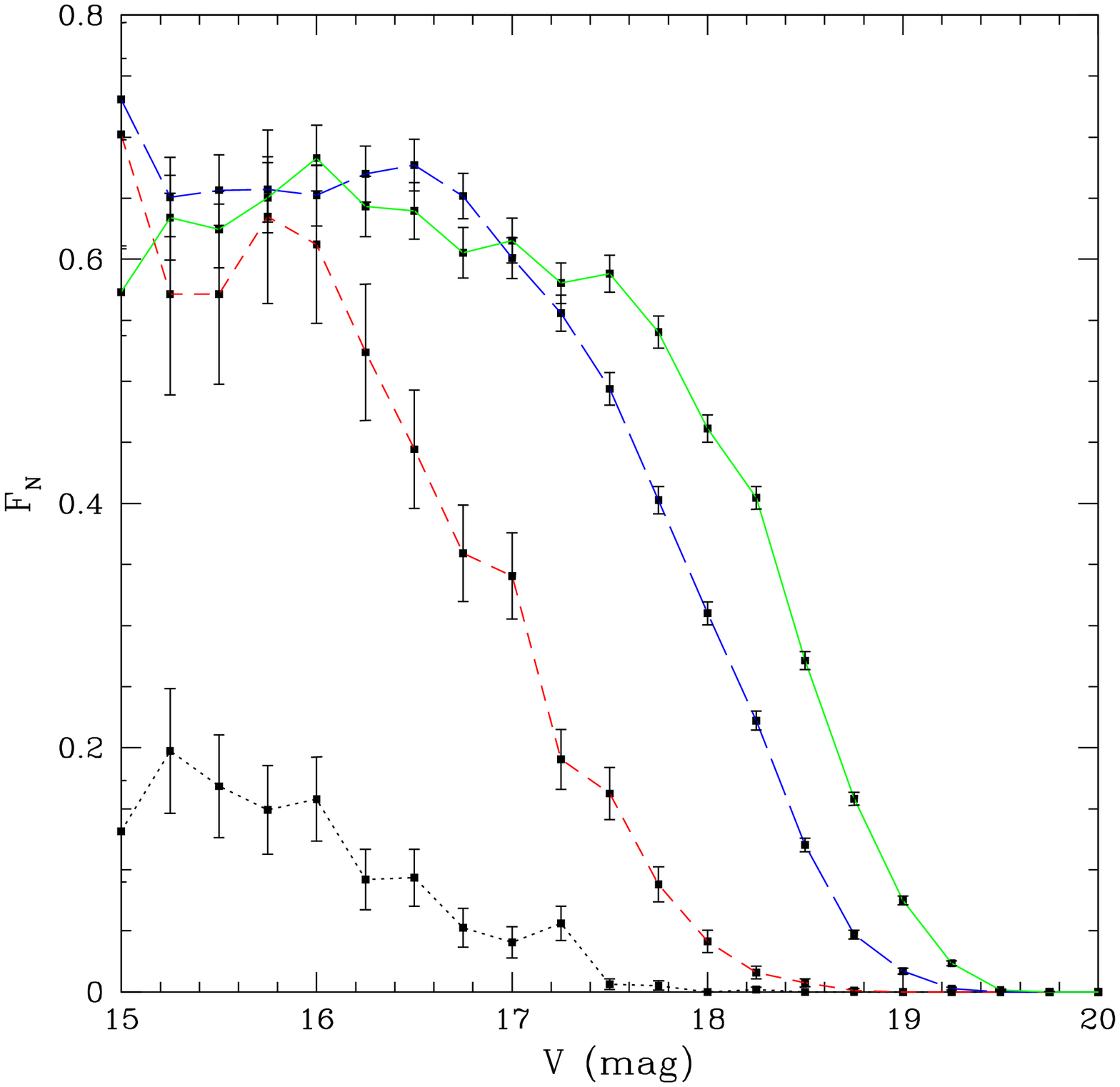}
\caption{\label{Compl} 
Detection completeness as a function of magnitude.
In the left figure the we show the fraction of all artificial lightcurves 
that were selected as variables as the solid line. The bashed-line 
shows the fraction after selecting object in the correct period range
and processing through AFD software.
In the right figure we show the detection sensitivity for vary numbers
of observations in a field. The dotted-line shows the result for fields 
sampled an average of 80 times, the short-dashed line for fields sampled 
115 times, the long-dashed line 160 times and the solid line 200 times.
}
}
\end{figure}

\begin{figure}{
\epsscale{0.6}
\plotone{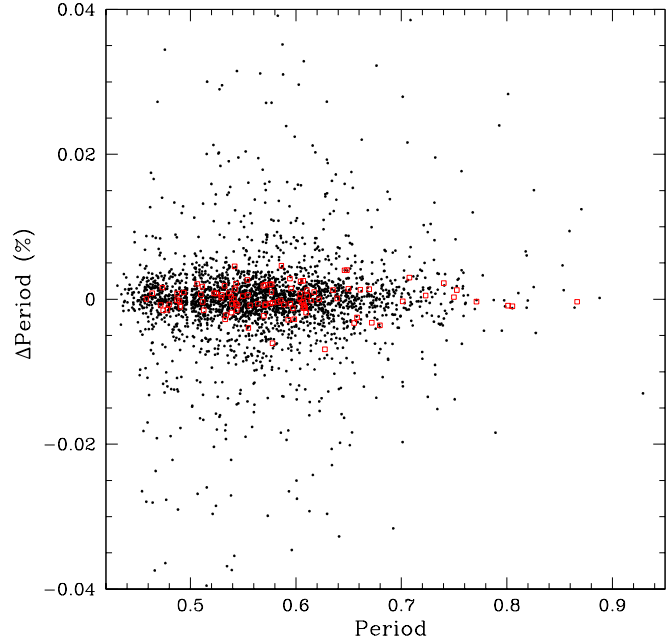}
\caption{\label{DiffVSX} 
Uncertainties in CSS RRab period determinations.
The black points show the percentage period difference between VSX and CSS periods for 
previously known RRab's. The boxes show the period differences for RRab with periods
determined separately in two overlapping CSS fields. 
}
}
\end{figure}

\begin{figure}{
\epsscale{1.0}
\plottwo{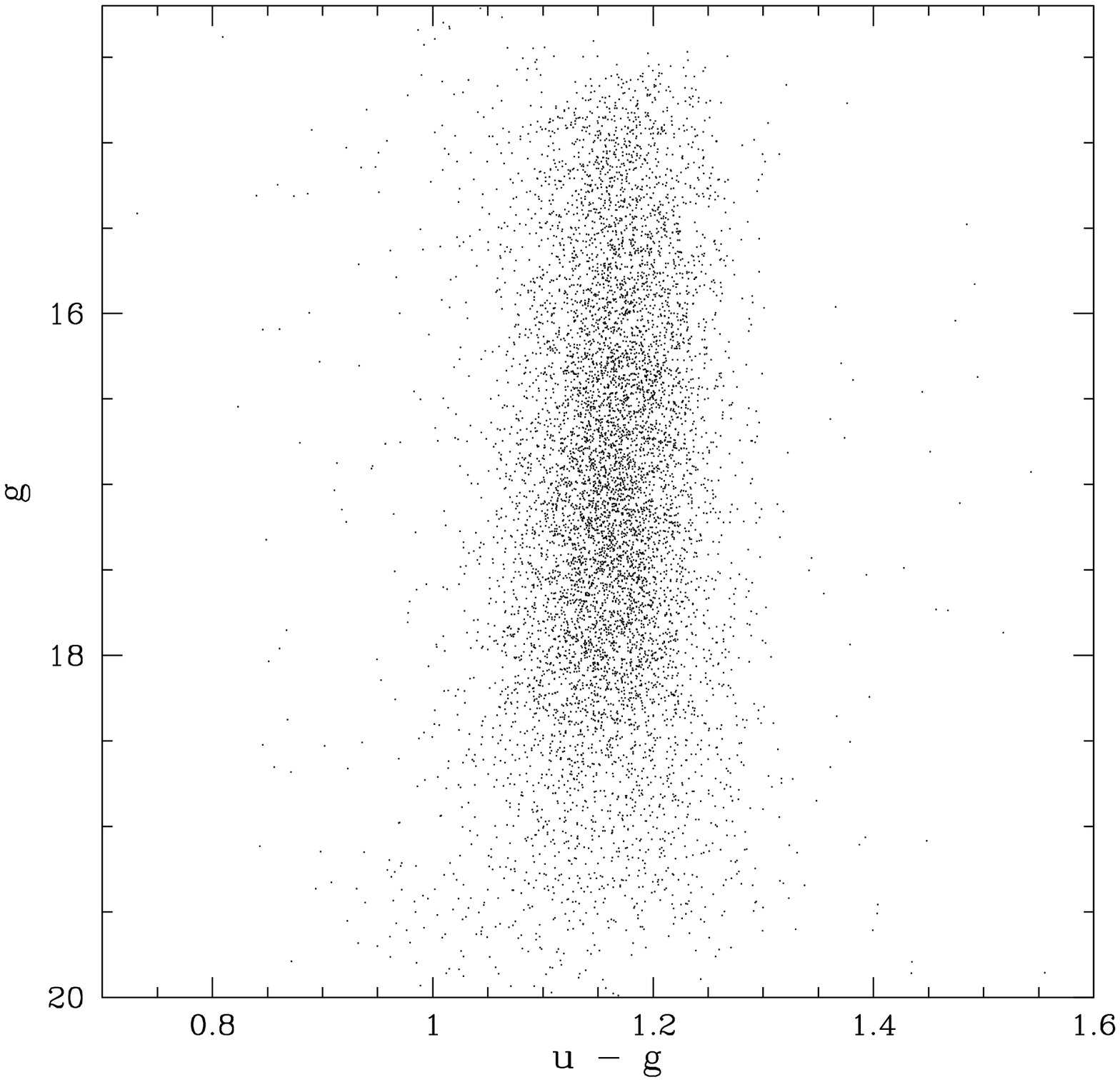}{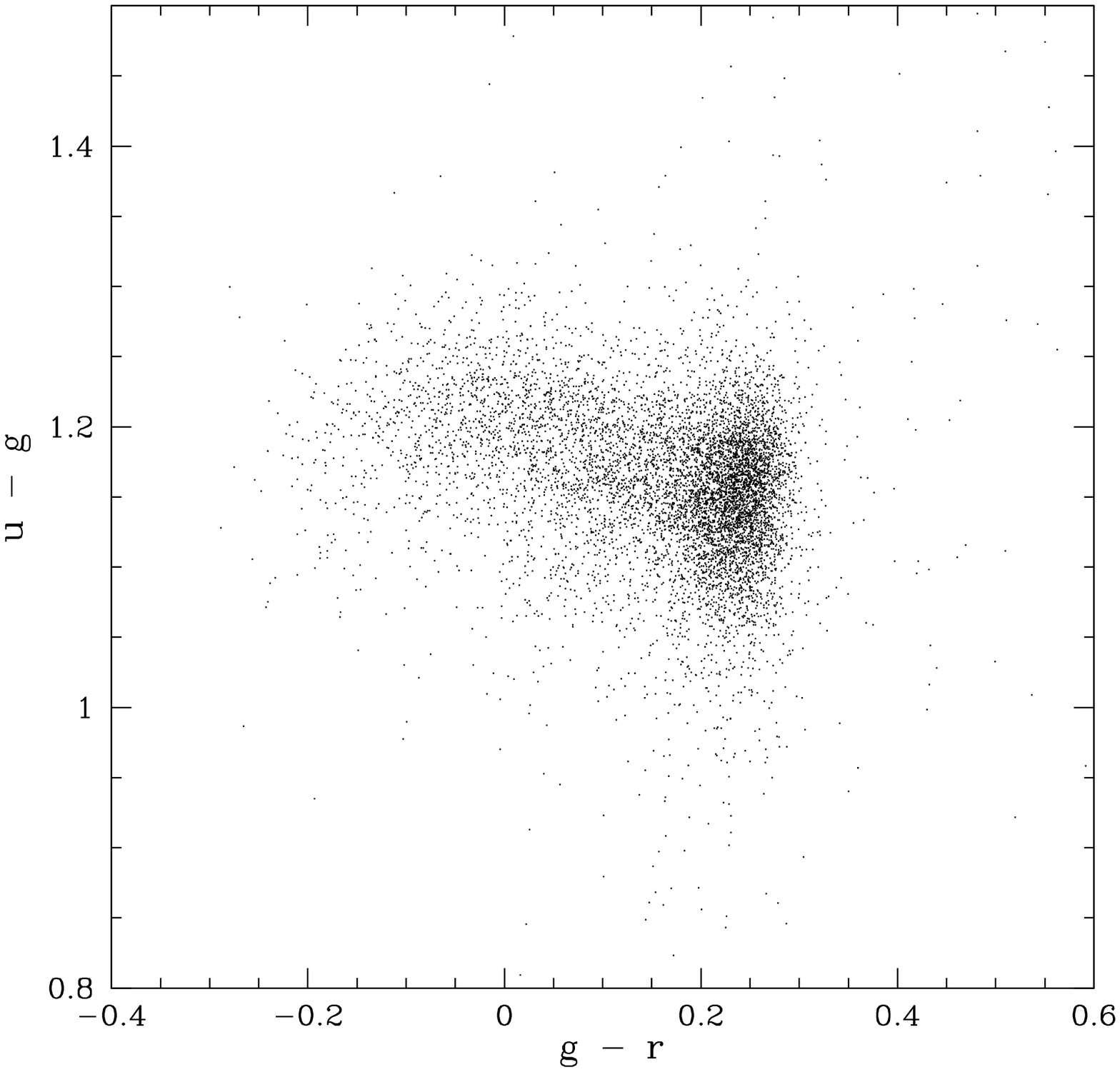}
\caption{\label{sdss}
Extinction corrected SDSS photometry for CSS RRab. Left: the 
colour-magnitude distribution of the RRL. Right: 
the $g-r$ vs $u-g$ distribution of the RRL.
}
}
\end{figure}

\begin{figure}{
\epsscale{1.0}
\plottwo{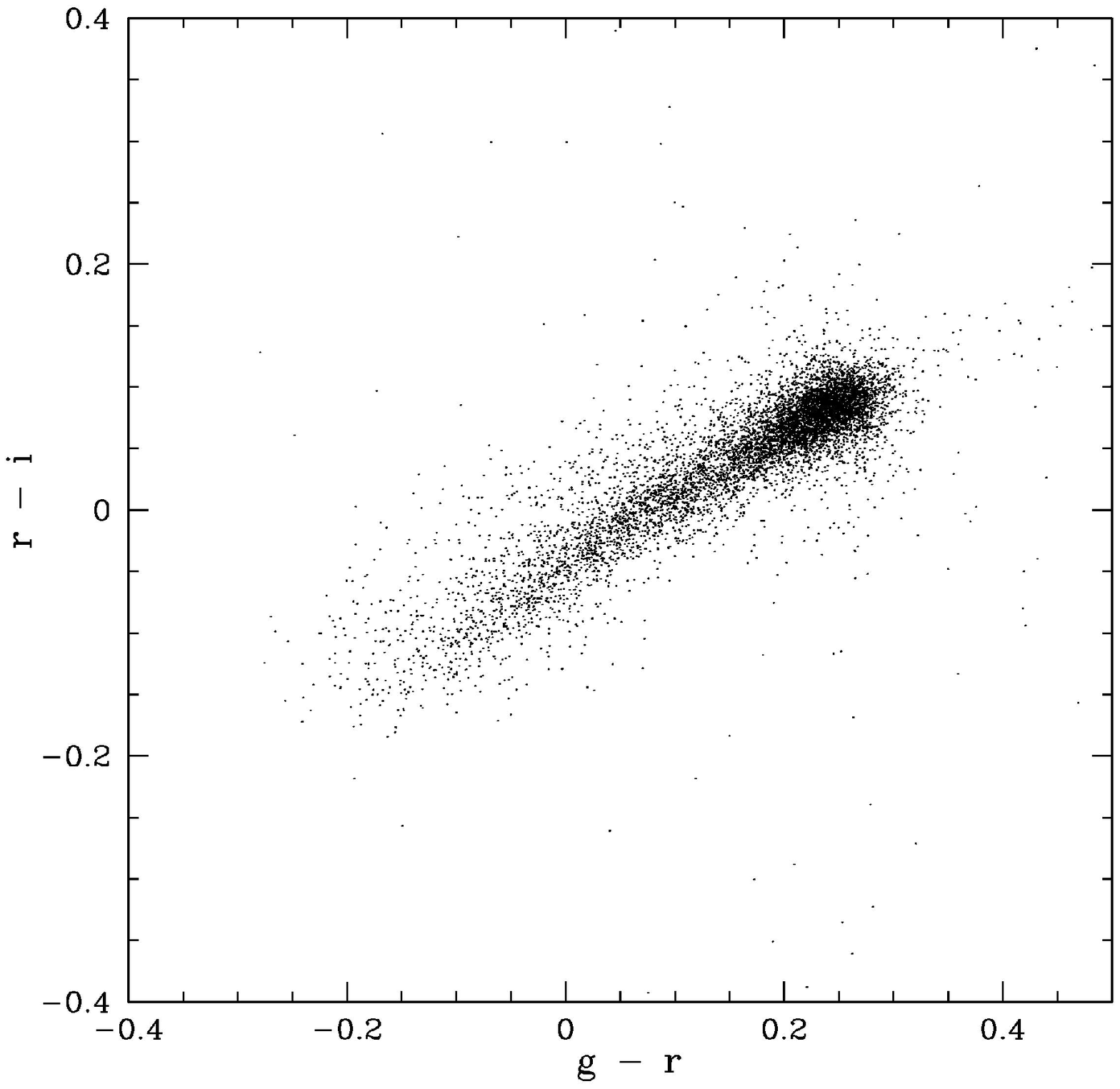}{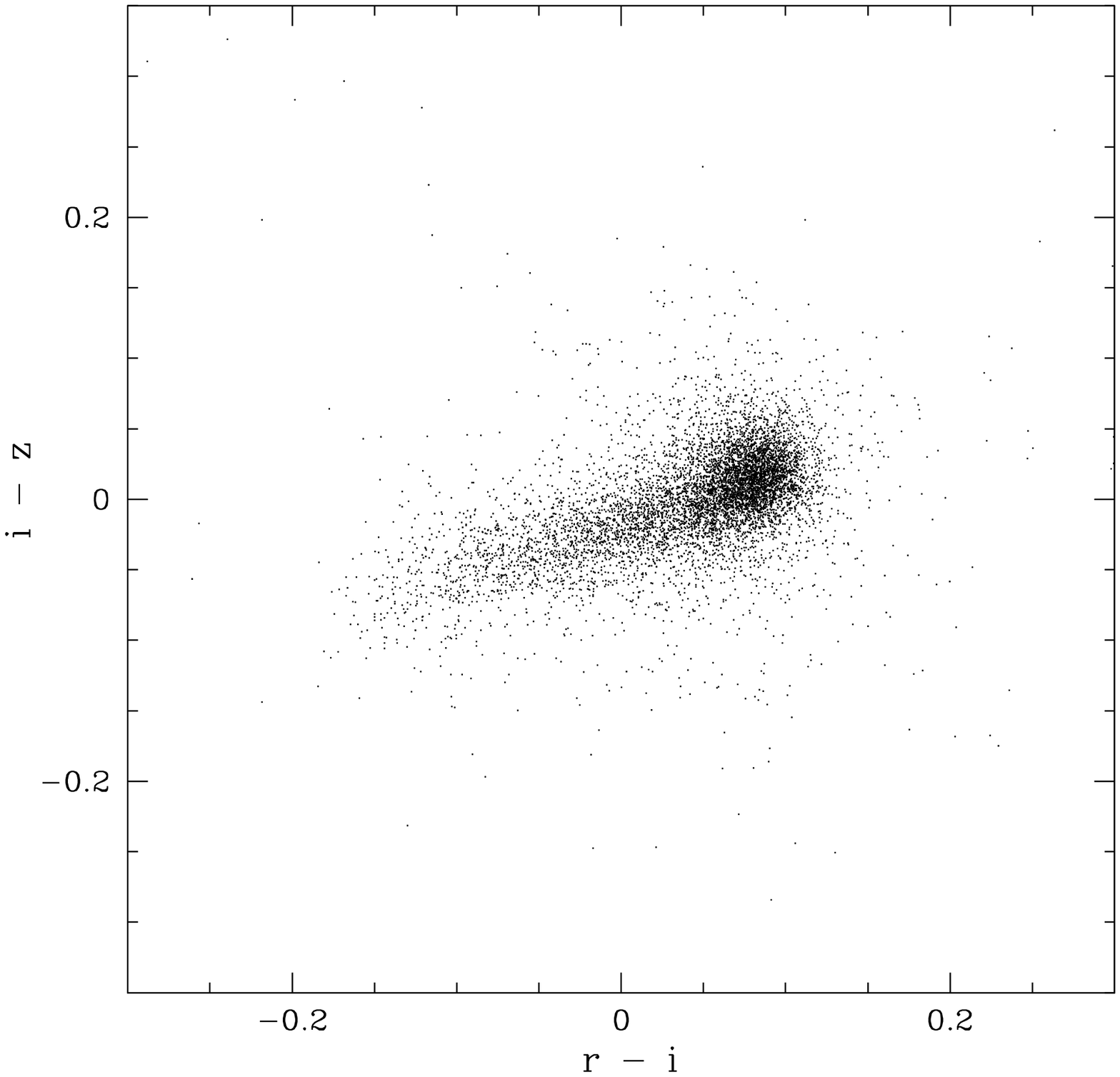}
\caption{\label{sdss2} 
Extinction corrected colour-colour plots for all CSS 
RRab with SDSS DR8 photometry.
Left: SDSS $g-r$ vs $r-i$ distribution of the RRL.
Right: SDSS $r-i$ vs $i-z$ distribution of the RRL.
}
}
\end{figure}

\begin{figure}{
\epsscale{1.0}
\plottwo{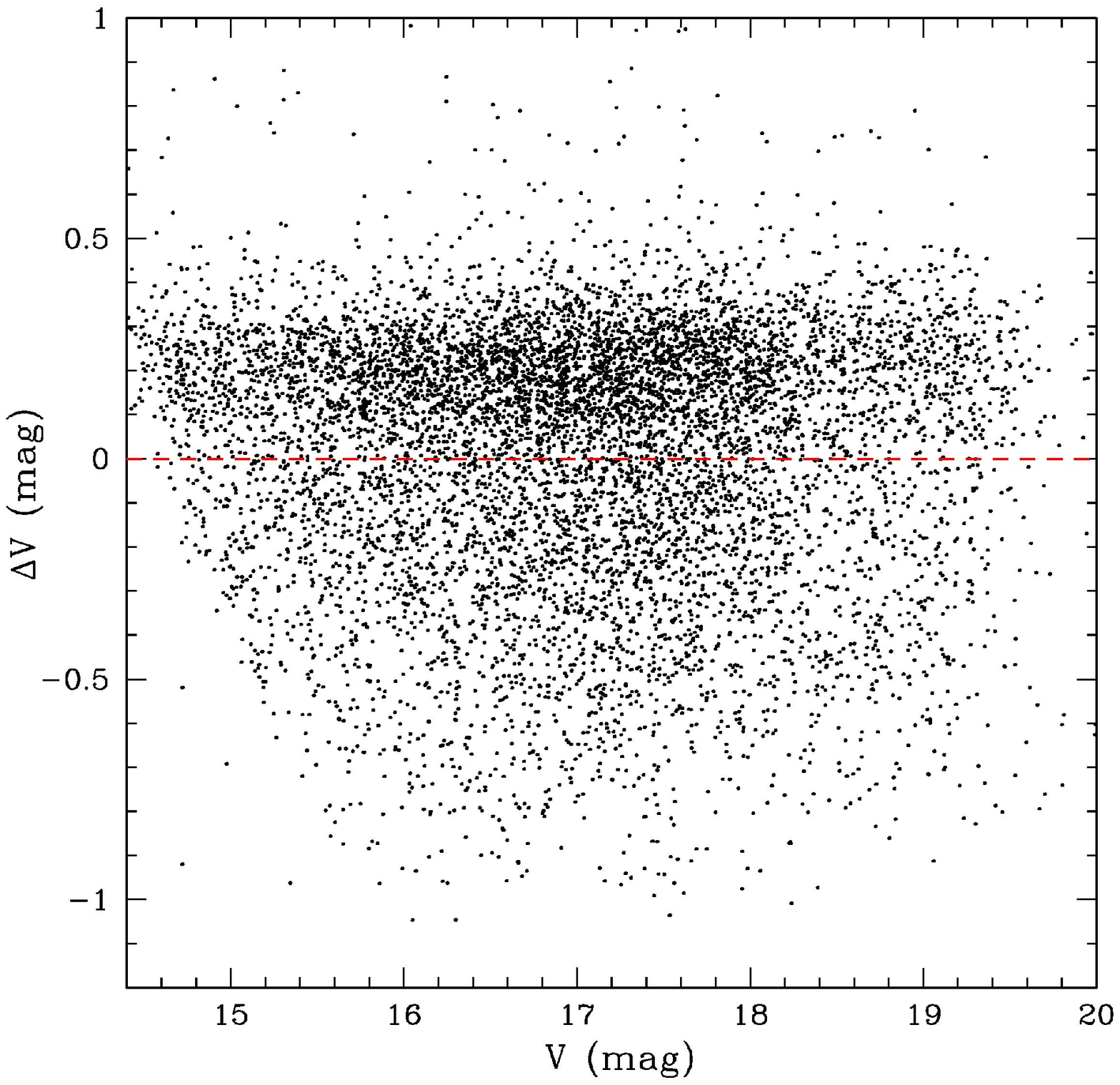}{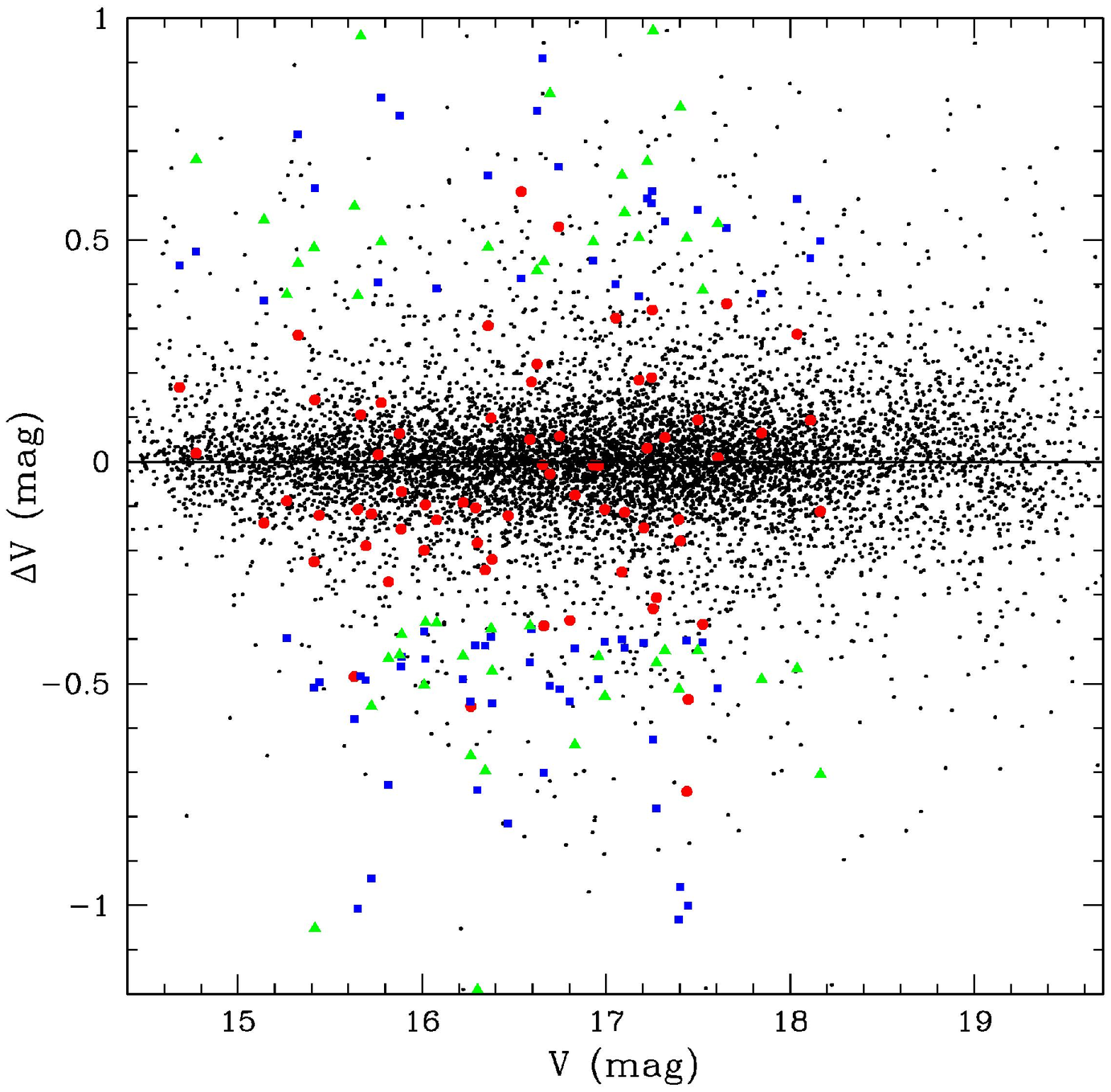}
\caption{\label{phase}
Difference between CSS and SDSS transformed magnitudes.
Left: the difference between SDSS and CSS magnitudes
transformed to V compared without correction for 
observation phase.
Right: The distribution of V-mag differences after
correcting for SDSS observation phase. The green
points show likely period changing RRab. The blue points
show differences for bright outlier sources before 
periods were refined. The red points show the difference 
for the same sources with improved periods.
}
}
\end{figure}

\begin{figure}{
\epsscale{0.6}
\plotone{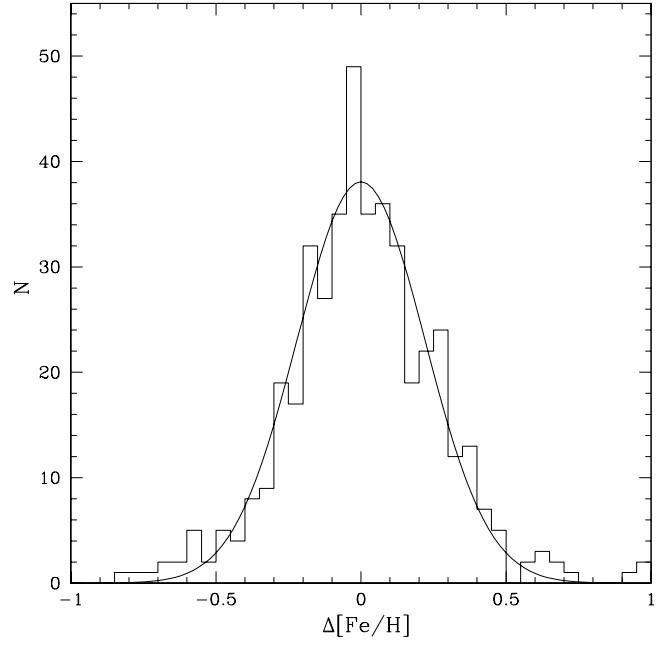}
\caption{\label{FeMult}
The distribution of differences in [Fe/H] calculated for RRab's
with multiple SDSS observations. The curve shows a Gaussian 
fit to the data.
}
}
\end{figure}

\begin{figure}{
\epsscale{0.6}
\plotone{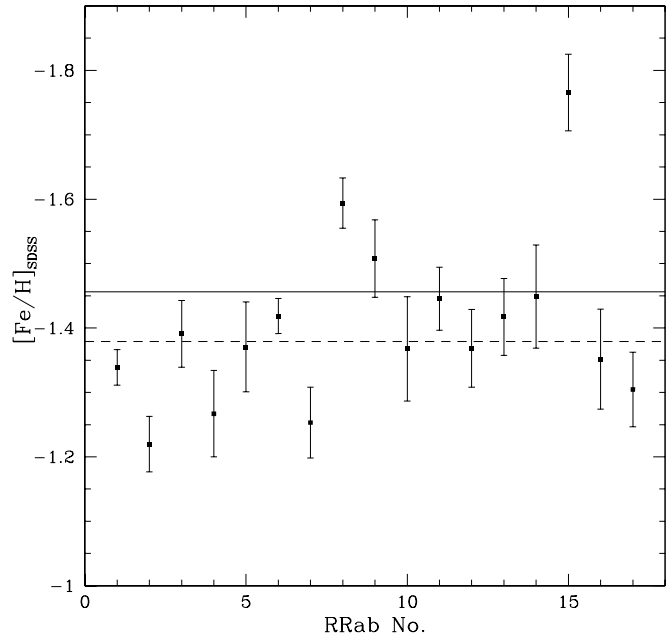}
\caption{\label{M3}
The distribution of metallicities for RRab's in NGC 5272
based on SDSS DR8 spectra. The solid-line shows the average 
value while the dashed-line shows the average from SDSS.
}
}
\end{figure}

\begin{figure}{
\epsscale{0.6}
\plotone{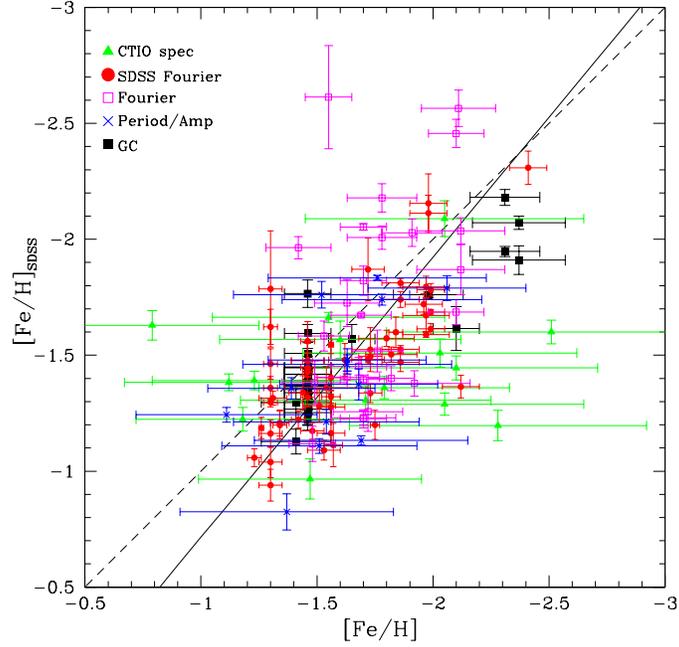}
\caption{\label{Fedelee}
Comparison between RRab metallicities derived by De Lee (2008)
and SDSS DR8 values. Twenty six additional globular cluster RRab's 
with SDSS spectra have been included. Each point is marked with
a symbol presenting the method used to determine the value as noted
in the text. A straight-line fit to the data is given by the 
solid-line, whereas a the dashed-line shows the slope assuming
no difference between the SDSS and De Lee (2008) values.
}
}
\end{figure}

\begin{figure}{
\epsscale{0.6}
\plotone{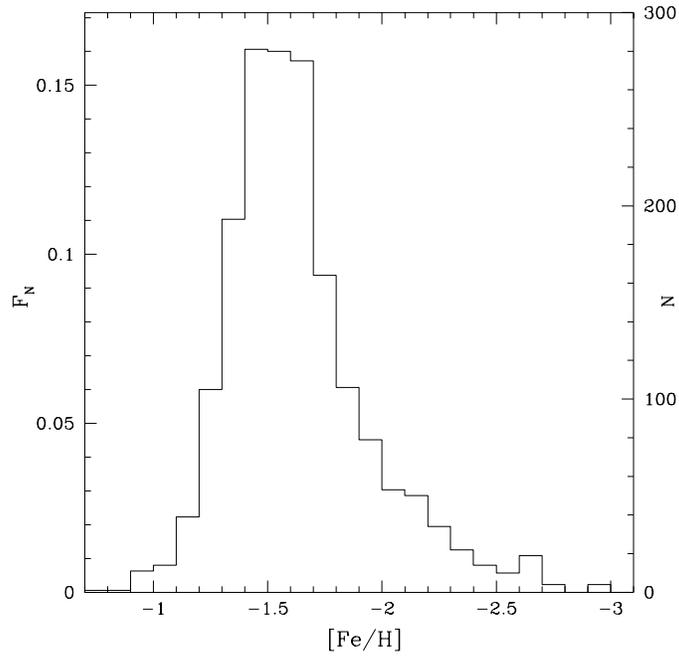}
\caption{\label{FE}
The metallicity distribution of 1382 RRab's in the CSS sample from 
1749 SDSS spectra with metallicity measurements. Fractional numbers 
of RRab, $\rm F_N$, are plotted on the ordinate axis.
}
}
\end{figure}

\begin{figure}{
\epsscale{0.6}
\plotone{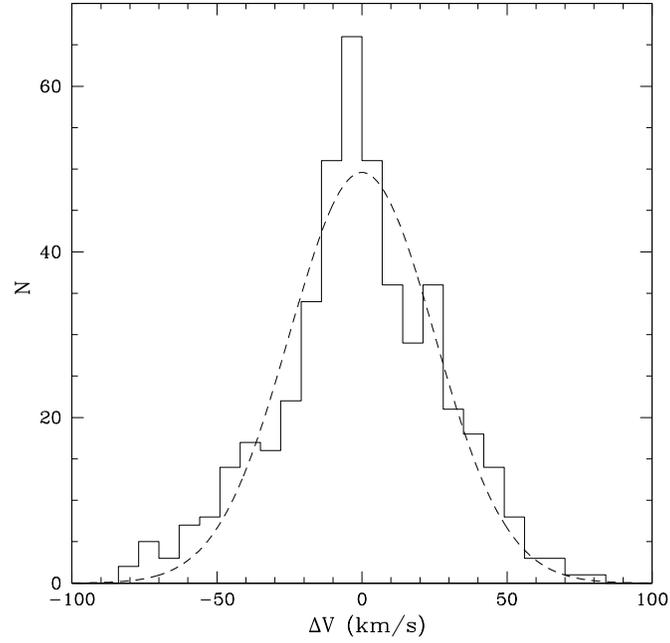}
\caption{\label{RadPul}
The distribution of differences in RRab radial velocities calculated
from pairs of SDSS spectra. The dash-line presents a Gaussian
fit to the distribution.
}
}
\end{figure}

\begin{figure}{
\epsscale{0.6}
\plotone{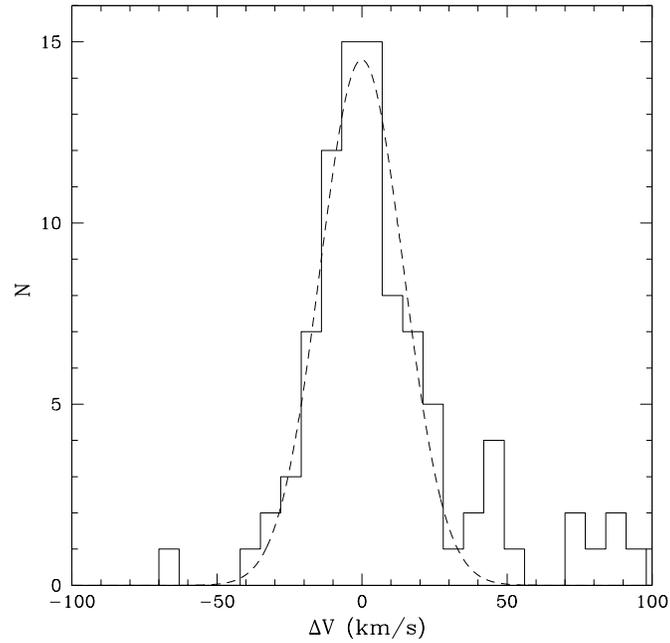}
\caption{\label{RadPulcor}
The distribution of differences in RRab radial velocities calculated
from pairs of SDSS spectra after corrections have been made for
pulsational velocity. The dash-line presents a Gaussian fit 
to the distribution.
}
}
\end{figure}

\begin{figure}{
\epsscale{0.6}
\plotone{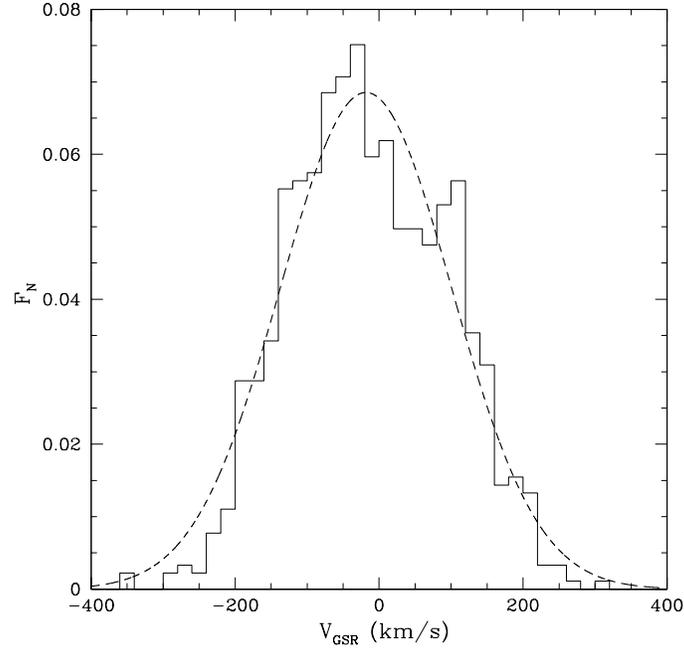}
\caption{\label{Vel}
The distribution of RRab radial velocities relative to the
Galactic standard of rest ($\rm V_{GSR}$). This histogram 
contains values for the 905 CSS RRab with SDSS DR8 radial 
velocity measurements.  The Gaussian fit the distribution 
is also plotted. The plot is presents the fractional number 
of objects, $\rm F_N$. 
}
}
\end{figure}

\begin{figure}{
\epsscale{1.0}
\plottwo{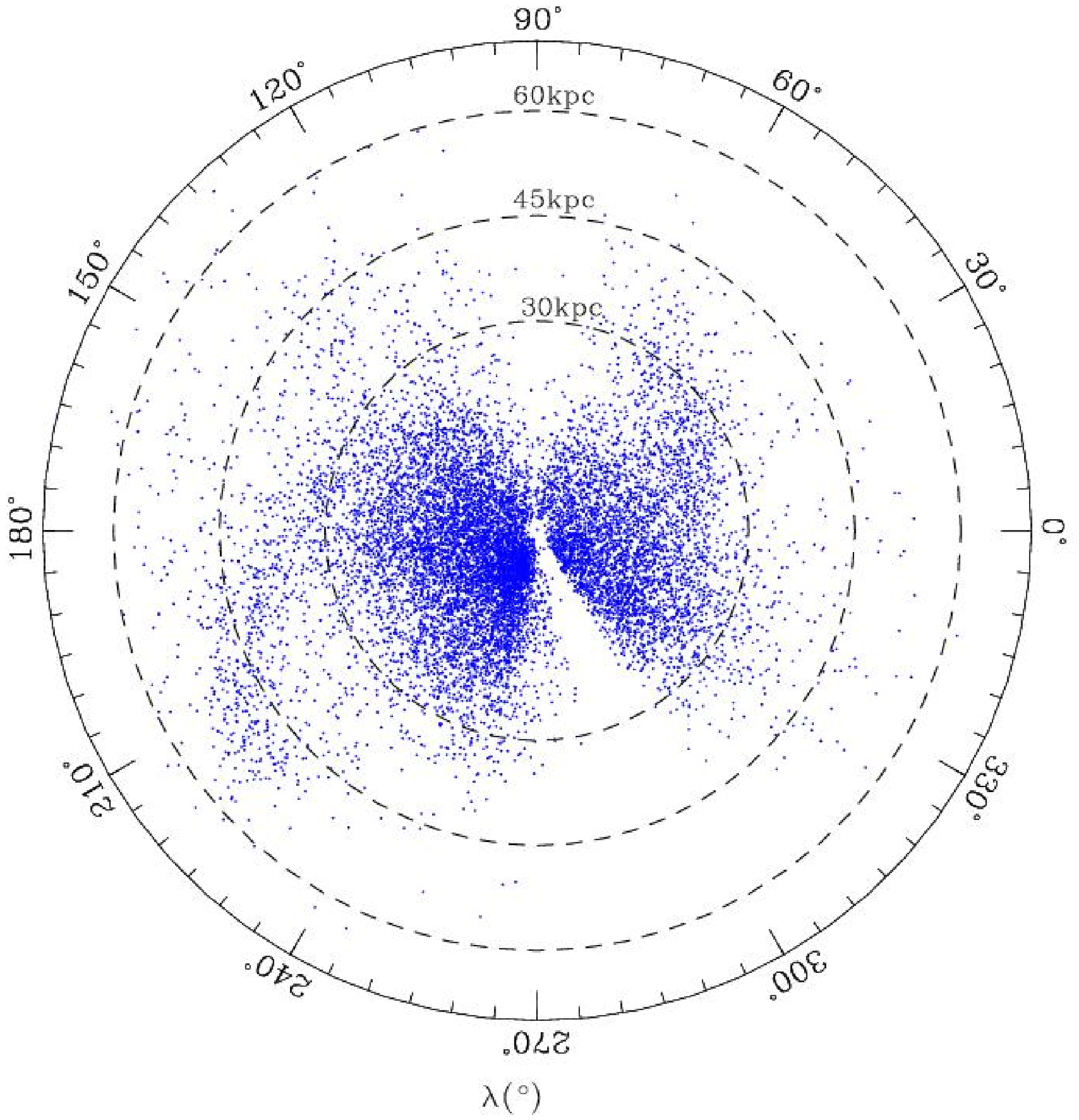}{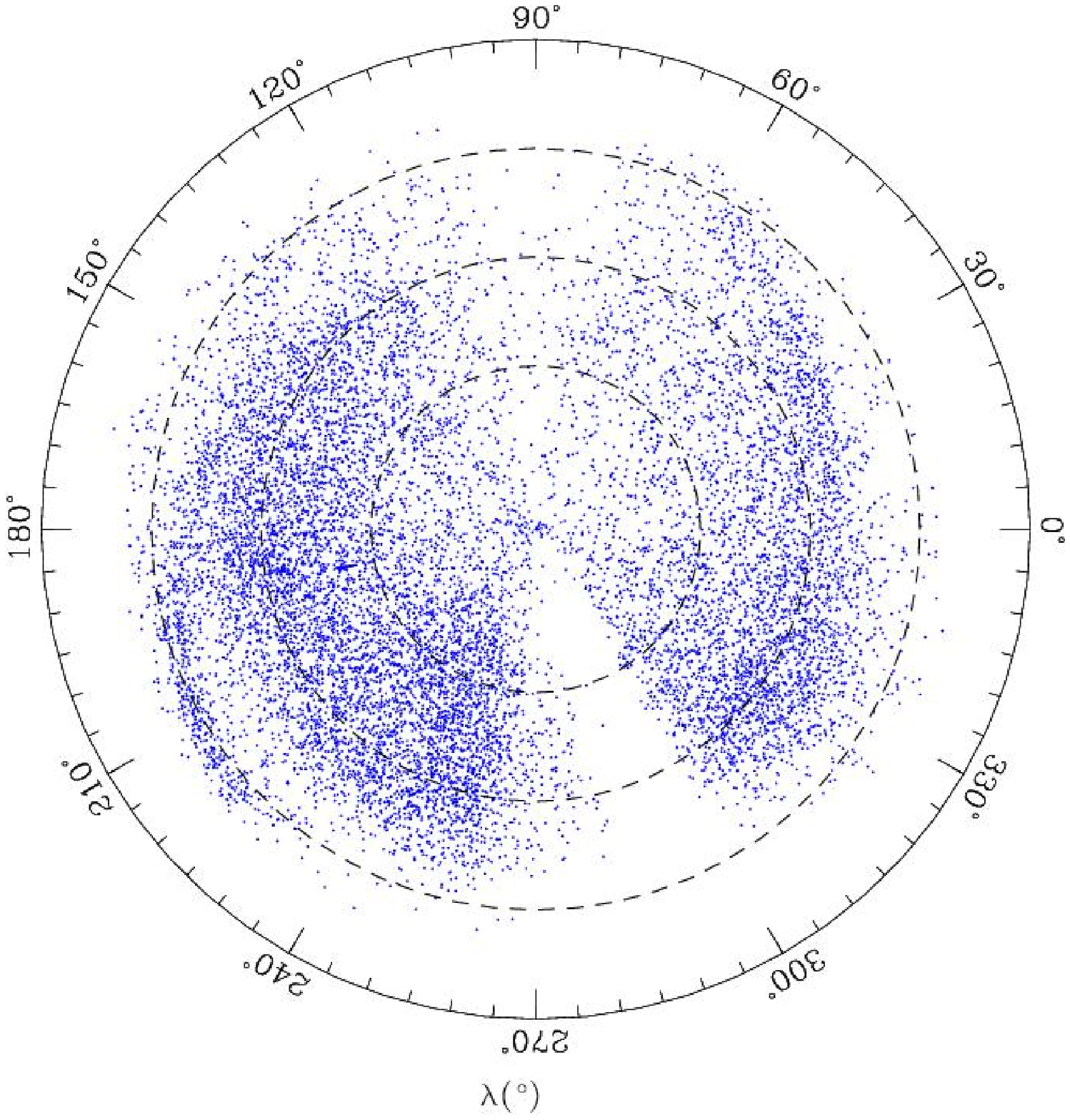}
\caption{\label{Eclip}
The magnitude and distributions of CSS RRab in ecliptic coordinates.
The left plot gives the distribution of RRab heliocentric distances. 
The right plot gives the distribution of magnitudes for RRab with 
$(V_0)_S > 14$. Here the dashed lines are set at magnitudes 
$V=15$, 17 and 19.
}
}
\end{figure}
\clearpage

\begin{figure}{
\epsscale{0.65}
\plotone{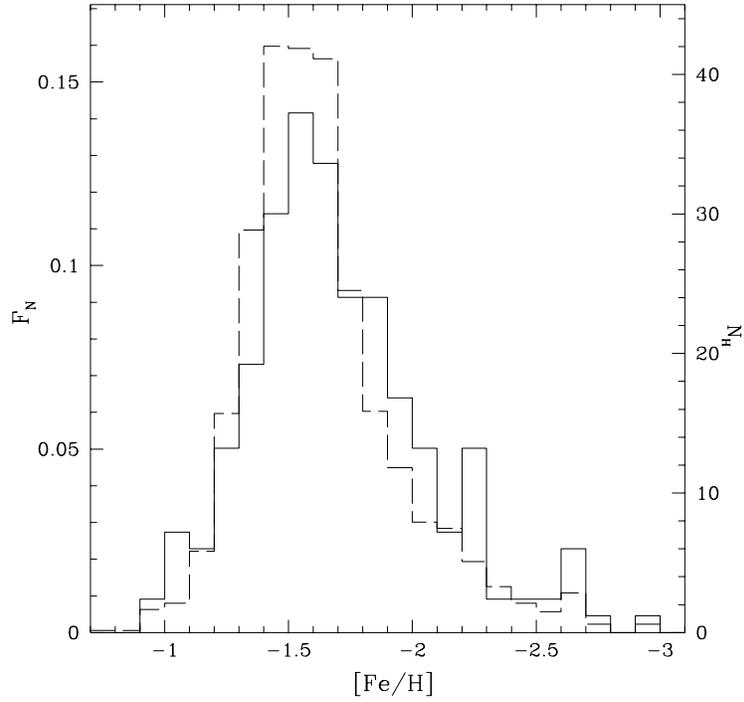}
\caption{\label{FeSgr}
The distribution of SDSS metallicities for Galactic halo RRab's 
with Galactocentric distances $r_{G} > 33.5$ kpc. Here $\rm F_N$
is the fractional number of RRab and $N_H$ the actual number
of Halo RRAb's. The solid-line gives the halo RRab's and the 
dashed line give the distribution for all CSS RRab's with SDSS 
metallicities.
}
}
\end{figure}

\begin{figure}{
\epsscale{0.9}
\plotone{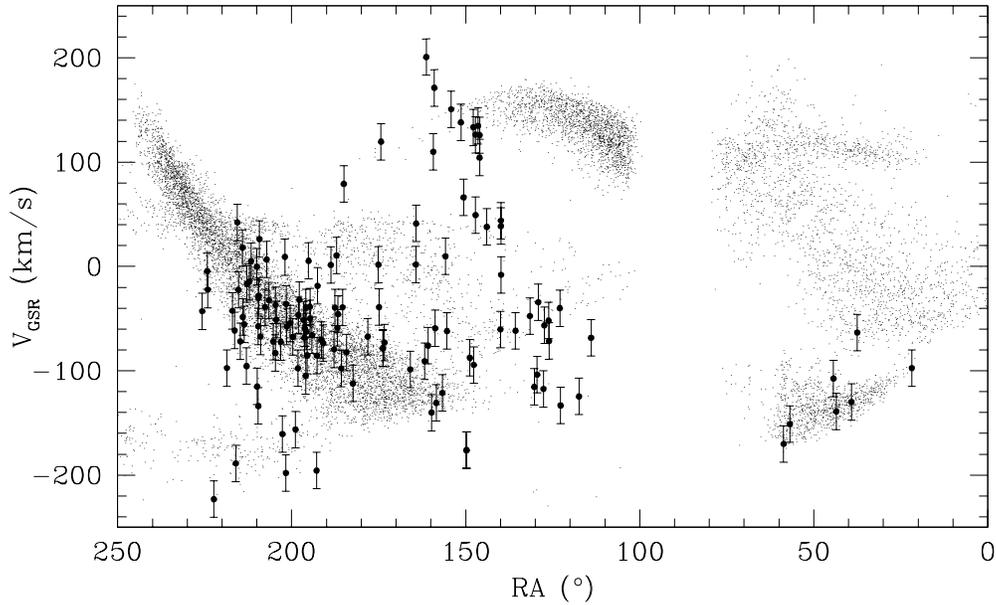}
\caption{\label{VelSgr}
The distribution of Halo RRab radial velocities for possible Sgr stream members. 
The large dots show the SDSS radial velocities relative to the Galactic standard
of rest for CSS RRab's with $d_h > 30$ kpc, and $-15\arcdeg < B < -15\arcdeg$ 
in Sgr stream coordinate system (Majewski et al.~2003). In addition, the velocities 
are plot for LM10 model data points within the area and distance range covered
by the CSS data.
}
}
\end{figure}

\begin{figure}{
\epsscale{0.8}
\plotone{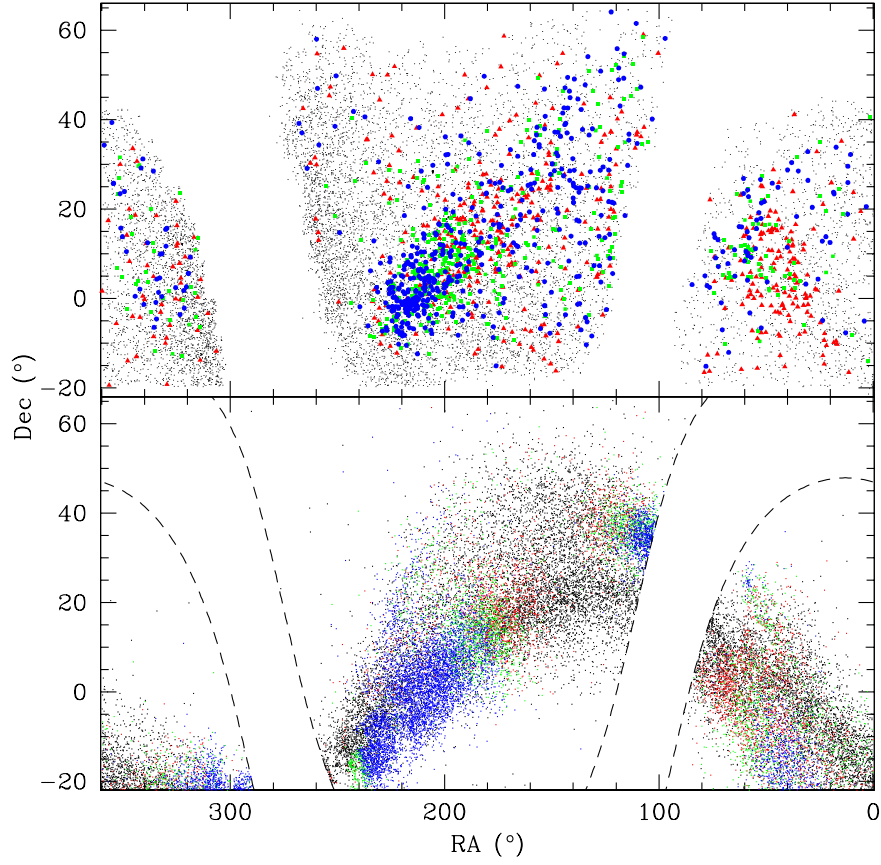}
\caption{\label{CompLaw}
The spatial distribution of CSS RRab's.
In the top panel we plot CSS RRab's at Galactocentric distances
$< 33.5$ kpc as black points, 33.5 to 38 kpc as red triangles, 
38 to 44 kpc as green squares  44 to 65 kpc as blue circles.
In the bottom panel we plot points from the LM10 model using 
the same colours.
}
}
\end{figure}

\begin{figure}{
\epsscale{0.6}
\plotone{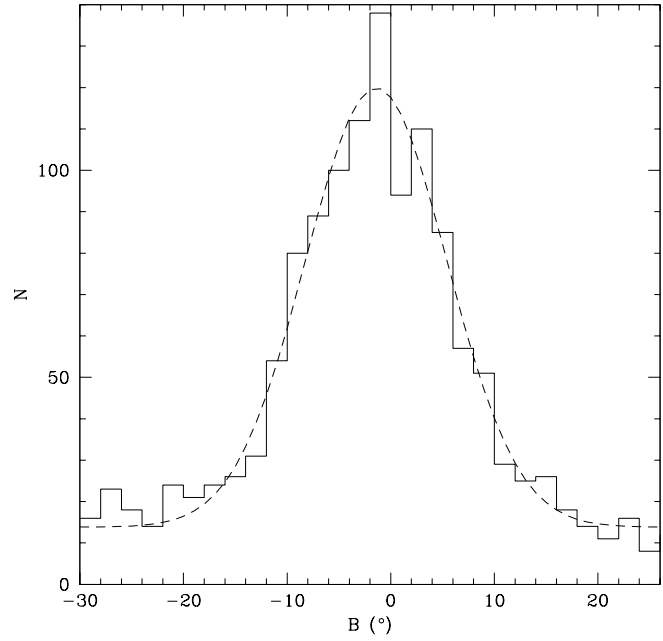}
\caption{\label{DistSgr}
The density distribution of RRab's with $d_h > 30$ kpc in the 
Majewski et al.~(2003) Sagittarius stream coordinate system.
The dashed-line presents a Gaussian fit to the data.
}
}
\end{figure}

\begin{figure}{
\epsscale{0.7}
\plotone{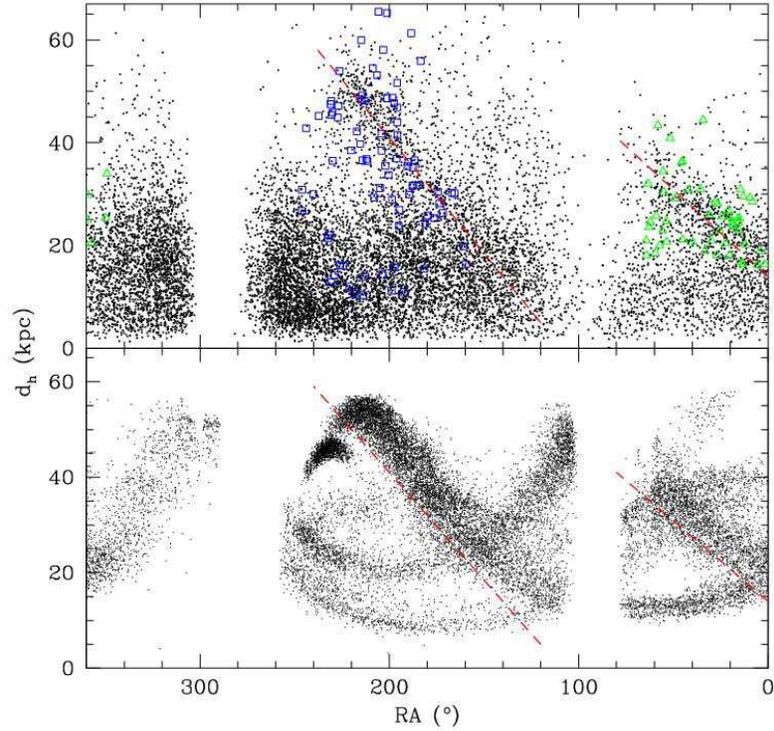}
\caption{\label{DistRA}
The distribution of RRL distances within the Sgr streams region. Here we have selected 
RRL and simulated objects in the Sagittarius coordinate system (Majewski et al.~2003) with 
$-11\arcdeg < B < 11\arcdeg$. In the top panel we plot RRab's as well as leading-stream M-giants
(squares) and trailing-stream M-giants (triangles) from LM10. In the lower panel, we plot the 
LM10 model truncated at 60kpc after taking into account the CSS survey spatial coverage and extinction.
The dashed-lines represent the estimates of mean distances of RRab in leading and trailing Sgr arms. 
These are de1fined as $d_h =0.45\times\delta - 49$ (kpc) and $d_h =0.338\times\delta + 14$ (kpc), 
respectively. A higher resolution figure is available in the online journal.
}
}
\end{figure}

\end{document}